\documentclass[preprint,12pt]{elsarticle}

\usepackage{epsfig}

\usepackage{amssymb}

\begin{document}

\begin{frontmatter}
\title{{\small \emph{Advances in Adaptive Data Analysis}, Vol.5, No.3 (2013), 1350015 (36 pages).}\\[-1mm]
{\small [DOI:10.1142/S1793536913500155, World Scientific Publishing Company]}\\[3mm]
{Extreme-Point Symmetric Mode Decomposition Method for Data Analysis}}


\author{Jin-Liang Wang\footnote{Corresponding author.}, \quad Zong-Jun Li}

\address{{College of science, Qingdao Technological University,}\\
{East Jialingjiang Road No.777, Huangdao Region of Qingdao, 266520, P.R. China.}\\
 {E-mail: wangjinliang0811@126.com (J. L. Wang),\enskip  li\_zjun@126.com (Z. J. Li)}}

\baselineskip = 18 pt

\begin{abstract}
\quad\enskip \small  An extreme-point symmetric mode decomposition (ESMD) method is proposed to improve the Hilbert-Huang Transform (HHT) through the following prospects: (1) The sifting process is implemented by the aid of 1, 2, 3 or more inner interpolating curves, which classifies the methods into ESMD\_I, ESMD\_II, ESMD\_III, and so on; (2) The last residual is defined as an optimal curve possessing a certain number of extreme points, instead of general trend with at most one extreme point, which allows the optimal sifting times and decompositions; (3) The extreme-point symmetry is applied instead of the envelop symmetry; (4) The data-based direct interpolating approach is developed to compute the instantaneous frequency and amplitude. One advantage of the ESMD method is to determine an optimal global mean curve in an adaptive way which is better than the common least-square method and running-mean approach; another one is to determine the instantaneous frequency and amplitude in a direct way which is better than the Hilbert-spectrum method. These will improve the adaptive analysis of the data from atmospheric and oceanic sciences, informatics, economics, ecology, medicine, seismology, and so on..
 \end{abstract}
\begin{keyword}
\small Extreme-point symmetric mode decomposition (ESMD); Empirical mode decomposition (EMD);
Hilbert-Huang transform (HHT); Direct interpolating (DI); Adaptive global mean (AGM);
Intrinsic mode function (IMF); Data (Signal) processing.
\end{keyword}

\end{frontmatter}



\baselineskip = 18 pt

\section{Introduction}
\quad\enskip Adaptive data analysis method plays a critical role in understanding the physical processes without mature mathematical models. In order to find the hidden laws underneath the rambling stochastic data,  the  common  way  is  to  decompose  it  into  a  series  of  modes  with  different frequencies. Its progress relies on the innovation of analyzing methods. Among all of the data analysis methods, the classical one is Fourier transform. It is based on linear superposition principles with a mapping from the time series to the space of frequency-energy spectrum, which accords with infinite modes in the form of sinusoidal functions with fixed amplitude and frequency. Therefore, it is only suitable for the stationary cases associated with linear problems. The Wavelet transform is very popular nowadays. It decomposes the data into a series of finite modes by aid of a series of frequency windows, which can provide a time-varying frequency for  non-stationary time series, however, its theoretical basis is also the linear superposition principle. To overcome these difficulties, Huang \emph{et al.} (1998) developed an adaptive method, named ``Hilbert-Huang transform (HHT)", which is very hot nowadays. It requires neither priori primary-function nor preset window-length. For this case, the linear superposition principle is also abandoned. These characteristics allows the better study of the non-stationary time series associated with nonlinear problems [Huang and Shen (2005)].

 \vskip 1mm There are two parts for HHT method. The first and the key part is called ``empirical mode decomposition (EMD)" which yields a series of intrinsic mode functions (IMFs), the second part
is called ``Hilbert spectral analysis (HSA)" which yields meaningful
instantaneous amplitude and frequency for these IMFs.
  The scheme of EMD is to make a mode symmetric about its upper
and lower envelopes interpolated by the local maxima and minima points separately.

\vskip 1mm By carrying forward the methodology of EMD, in this paper, we develop an ``extreme-point symmetric mode decomposition (ESMD)" method based on the natural phenomenon that a high-frequency small wave rides on a low-frequency big wave and the small one is almost symmetric about its crest and trough relative to the big one. Enlightened by this fact, we introduce a new scheme to make mode symmetry. Differing from constructing 2 outer envelopes, the sifting process is executed by the aid of 1, 2, 3 or more inner curves interpolated by the midpoints of the line segments connecting the local maxima and minima points. The similar approach  had been discussed by Huang \emph{et al.} (1998) for the 1-interpolating-curve case. However, it will be shown that 2 or 3 interpolating curves can lead better decompositions.

\vskip 1mm There are five important issues associated with the developments of ESMD:
the stoppage criteria; the global mean curve; the concepts of symmetry and periodicity;
 the instantaneous frequency and the definition of IMF.
 In the following we give some reviews on the related topics.

\subsection{About the Stoppage Criteria}

\quad\enskip How to choose the stoppage criteria remains to be an open problem, since different sifting times may result in different decompositions [Huang \emph{et al.} (2003), Huang and Wu (2008)]. On the one hand, less  times of sifting may lead poor symmetry to the IMFs and render inaccuracy to the analysis of the instantaneous frequency by Hilbert transform, but on the other hand, a large number of sifting is not recommended [Huang \emph{et al.} (2003), Wu and Huang (2009, 2010), Wang \emph{et al.} (2010)], since over-sifting probably obliterates the intrinsic amplitude variations and leads to unphysical results.
The theoretical study given by Wang \emph{et al.} (2010) indicated that the upper and lower envelopes (in form of cubic splines) of a rigid symmetric IMF with sparsely populated extreme points have to degenerate to a pair of symmetric straight lines. In the subsequent paper, Wu and Huang (2010) pointed out that the sparse condition is not necessary. These theoretical results lead to a conjecture that the equal-amplitude IMFs appear for a high enough sifting times. Though this conjecture is very attractive, it is out of reach for the actual sifting process. Our recent research indicates that the change of symmetry degree for IMFs is in an intermittent manner [Wang and Li (2012a)].
In another word, as the sifting times is added, the sustaining-modulation state and
sudden-turn state will appear alternatively. So the symmetry of IMF will become better for the case
of sustaining-modulation and worse for the case of sudden-turn.
In addition, it follows from sifting tests that, after a certain sifting times, the average frequency of each IMF keeps
steady or changes in an oscillating manner. Hence, the frequency ratio for the neighboring IMFs
can not decrease to $1$ and the conjecture given by Wu and Huang (2010) is cracked.
The dyadic filter bank property of EMD
[Flandrin, Rilling and Goncalves (2004), Flandrin and Goncalves (2004), Flandrin, Goncalves and Rilling (2005),
 Wu and Huang (2005)] and the frequency decomposition problem
 [Rilling and Flandrin (2008), Wu, Flandrin and Daubechies (2011)] can be also reconsidered for this limit case.

\vskip 1mm
In summary, there are four types of stoppage criteria [Wang \emph{et al.} (2010), Wang and Li (2012a)]:
(1) the Cauchy type [Huang \emph{et al.} (1998), Huang and Wu (2008)];
(2) the mean curve type [Rilling \emph{et al.} (2003), Wang and Li (2012a)];
(3)the $S$-number type [Huang \emph{et al.} (2003)];
(4) the fixed-sifting-times type [Wu and Huang (2009, 2010)].
 Among all these choices, if the mode symmetry is merely concerned in the sifting process then
the mean curve type criteria are preferable, after all, the symmetry degree of IMFs changes in an intermittent manner
along the sifting times. In this sense,
the fixed-sifting-times type of criteria are not preferred
if there is no prejudgement on the symmetry. With the help of the mean curve type criteria we
have developed an ensemble ``optimal-sifting-times" one for decomposition.

\subsection{About the Global Mean Curve}

\quad\enskip For the given data, its frequency analysis should be done on the oscillation part.
It  is the first and foremost problem to remove the global mean or trend.
It is noted  that  the total  mean (mathematical expectation in statistics) is just the
simplest form of the global mean curve.
To extract it, the common least-square method and running-mean approach are usually used.
The least-square method may provide an optimal fitting curve for a given data in the sense of
least variance. But it is awkward in application for the requirement of a priori function form.
The running-mean approach assigns the weighted mean value of several points to their center one,
it may provide a smooth global mean curve to a given data.
But this approach lacks of theoretical basis and
different choices of time-window and weight coefficients may result in
different curves. Physically speaking, some
process may change in a memory-dependent manner as indicated by
Wang and Li (2011) rather than a prospective-dependent.

\vskip 1mm Since the EMD method adopts an adaptive scheme, to some extent,
the global mean curve in form of trend function (with at most one extreme point) can be well extracted.
But this kind of global mean curve may miss the evolutionary trend
due to the bending limit. In order to make up this defect, it requires
an superimposing management on the lower-frequency modes.
Certainly, how to determine the number of the modes is a problem.
Moghtaderi and his partners
[Moghtaderi, Borgnat and Flandrin (2011), Moghtaderi, Flandrin and Borgnat (2011)]
have discussed it by using an energy-ratio approach.
Because this approach is involved in the ratio of zero-crossing numbers
(identical to that of frequency) between the neighboring IMFs which is sensitive to the sifting times [Wang and Li (2012a)],
whether the obtained global mean curve is optimal or not is still a problem.
In fact, rather than decomposing the data to the last trend function and
superimposing the lower-frequency modes in return, we can directly stop the decomposition
in a middle course.

\vskip 1mm
Differing from the EMD method, ESMD does not decompose the data to the last trend function with
at most one extreme point, instead, it permits the residual component possesses a certain number of extreme points.
One advantage of this processing is that: \emph{This kind of residual component can reflect the evolutionary trend
of the whole data much better and it can be understood as an optimal ``adaptive global mean (AGM)" curve};
The other advantage is that: \emph{We can optimize the
sifting times by optimizing this AGM curve in the least-square sense.}
In addition, this optimizing process itself is of great value.
It offers a good adaptive approach for data fitting which is superior to the common
least-square method and the running-mean approach.

\subsection{About the Concepts of Symmetry and Periodicity}

\quad\enskip The type of symmetry is directly related to understanding the concepts of periodicity.
Differing from the envelop-symmetry adopted by the EMD method, an extreme-point symmetry is adopted.
This difference makes us rethinking the fundamental concepts of periodicity.

\vskip 1mm
For a constant function or a monotone function, its periodicity and frequency arguments are not meaningful.
Only when a quantity varies in a periodic oscillating manner, the frequency can be
understood as an oscillating change rate.
The typical oscillating function is $A\cos{\omega t}$
which accords with the ideal periodic variation of the substance in nature.
Here $A$ and $\omega$ are called the amplitude and frequency.
But it is not always the case, such as a familiar damp vibration.
Though its frequency (determined by the material property) maintains unchanged, its amplitude
decreases as time goes on due to the air resistance.
 Certainly, the vibrating amplitude may also increase
if it gains energy. This phenomenon is very universal.
In order to describe it in mathematics, the concept of ``weighted periodicity"
is introduced in our previous works [Wang and Li (2006, 2007), Wang and Zhang (2006)].
A weighted-periodic function is the one with fixed frequency and varying amplitude in the form $A(t)\cos{\omega t}$.
Mathematically speaking, the concept of periodicity can be also enlarged to the general form
 $A(t)\cos{\theta(t)}$, where $\theta(t)$ is a continuous function. For convenience,
 we call it ``generalized periodic function".
 In fact, the aim of EMD sifting is to extract a series of IMFs of this form.
  Now that the IMFs are generalized periodic functions, they can be
 directly extracted by a mathematical approach. In this way,
 Hou and his partners [Hou, Yan and Wu (2009), Hou and Shi (2011, 2012)]
 have already made some effective explorations.

\vskip 1mm
 For a periodic function, since its amplitude and frequency are all fixed constants,
  its symmetric characteristic is very clear.
 It can be understood as the envelop symmetry or the extreme-point symmetry. But from the viewpoint of
 material movement, as a matter of fact, the oscillation occurs around the equilibrium position.
 So the extreme-point symmetry actually reflects the local symmetry about itself
 (all the midpoints of the line segments between the local maxima and minima points lie on the zero line).
 For a weighted periodic function, its frequency is fixed but its amplitude changes.
 For this case, its local symmetric characteristic is unclear, though
 it may appear envelop-symmetric on the whole. From the viewpoint of extreme-point symmetry,
 the equilibrium should also shift its location. Hence,
 \begin{equation}
 A(t)\cos{\omega t}=[A_r(t)+A_e(t)]\cos{\omega t},
 \end{equation}
 in which $A_r(t)$ and $A_e(t)$ should be understood as the amplitudes of real oscillating and equilibrium's shifting.
That means during the oscillating process the corresponding equilibrium also shifts its location
in the same frequency (see Fig.2). In addition, $A_r(t)$ and $A_e(t)$ are not independent,
after all, $A_e(t)\cos{\omega t}$ reflects the trajectory variation of the midpoint for $A_r(t)\cos{\omega t}$.
As for the generalized periodic function, not only the amplitude but also the frequency changes.
At this time, its amplitude can be also understood as above. But the frequency
can not be understood as a simple instantaneous one, after all, the shifting of
equilibrium location may distort the real oscillating frequency.
 For this case, the total oscillation
can be seen as a synthesis of two components:
\begin{equation}
A(t)\cos{\theta(t)}=A_r(t)\cos{\theta_r(t)}+A_e(t)\cos{\theta_e(t)}.
\end{equation}
Particularly, in case $A_e(t)\equiv 0$ the real amplitude $A_r(t)$ would degenerate to a constant and
the function becomes an equal-amplitude form $A_r\cos{\theta_r(t)}$.

 \vskip 1mm
This understanding is helpful for revealing the underneath nonlinear mechanism of a complex system.
The shifting phenomenon of equilibrium location is probably caused by the interaction
of vibrations with different frequencies. Corresponding to the mode decomposition,
this can be reflected by the interaction of IMFs. \textsl{It is the first attractive topic of this paper left for further discussion.}

\vskip 1mm
 According to the number of interpolating curves, we classify ESMD into ESMD\_I, ESMD\_II, ESMD\_III, $\cdots$.
 In fact, their differences lie in the request on the equilibrium variation $A_e(t)\cos{\theta_e(t)}$
which should be understood as an interpolating function with all the midpoints.
 ESMD\_I adopts a rigid extreme-point symmetry in the sifting process which requires all the midpoints almost lying on
the zero line, that is, $A_e(t)\approx 0$. For this case, all the IMFs should
almost degenerate to the equal-amplitude form $A_r\cos{\theta_r(t)}$.
From the viewpoint of physics, this strategy is too rigorous.
ESMD\_II extends the concept of the extreme-point symmetry, which permits the location shifting of the midpoint in
such a manner: \textit{The trajectory variation $A_e(t)\cos{\theta_e(t)}$
should be envelop-symmetric about its odd and even interpolating curves}.
These two envelops differ from the common positive and negative outer envelops, after all,
they may change their signs alternately.
The sifting test results show that this odd-even type of extreme-point symmetry for ESMD\_II
is almost equivalent to the outer envelop symmetry for EMD (see Fig.17).
ESMD\_III gives a further extension to the concept of extreme-point symmetry.
\textit{It liberalizes the restriction on $A_e(t)\cos{\theta_e(t)}$
and only requires the sum of two interpolating curves to be symmetric with the third one.}
Certainly, the restriction on $A_e(t)\cos{\theta_e(t)}$ can be also liberalized in this way with more interpolating curves.

\subsection{About the Instantaneous Frequency}

The definition of the instantaneous frequency is a controversial issue [Huang \emph{et al.} (2009a)].
 As analyzed above, only when a quantity varies in a periodic oscillating manner, the frequency can be
understood as an oscillating change rate during the process of moving back and forth.
So there is no local meaning for frequency at a given point.
But as argued by Huang \emph{et al.} (2009a) and the references therein, there is indeed
a frequency modulating phenomenon. Therefore, to accord with the
generalized periodic function $A(t)\cos{\theta(t)}$ in mathematics,
 the derivative form $\omega(t)=d\theta/{dt}$ is recommended.
 To be physically meaningful, it requires $d\theta/{dt}\geq 0$. However, for a decomposed IMF with denotation $x(t)$,
 this calculation is not trivial. In order to solve this problem, Huang \emph{et al.} suggested the Hilbert transform
  which is popular used nowadays:
 \begin{equation}
 y(t)=H[x(t)]=\frac{1}{\pi}P\int_{-\infty}^{\infty}\frac{x(\tau)}{t-\tau}d\tau,
 \end{equation}
 in which $P$ indicates the principal value of the singular integral. With the
Hilbert transform, the analytic data is defined as
\begin{equation}
z(t)=x(t)+iy(t)=A(t)e^{i\theta(t)},
\end{equation}
where
\begin{equation}
A(t)=\sqrt{x^2(t)+y^2(t)},\qquad \theta(t)=\arctan\left(\frac{y(t)}{x(t)}\right).
\end{equation}
Here $A(t)$ is the instantaneous amplitude and $\theta(t)$ is the phase function. Furthermore, the
instantaneous frequency can be calculated by $\omega(t)=d\theta/{dt}$.

\vskip 1mm
Essentially, the Eqn.(3) defines the Hilbert transform as the convolution of $x(t)$ and $1/t$,
 therefore, it emphasizes the local properties of $x(t)$. In Eqn.(4) the polar coordinate expression further clarifies the
local nature of this representation: it is the best local fit of an amplitude and phase varying
trigonometric function to $z(t)$ [Huang and Shen (2005)].
In this sense, the Hilbert transform is superior to the Fourier transform, Wavelet transform and other analytical forms.
However, this approach has a disadvantage in unsatisfying the quadrature request
 delimited by the well-known Bedrosian and Nuttall theorems.
 That is, in order to use Eqn.(4) $y(t)$ should be a quadrature function of $x(t)$.
 In addition, to get a meaningful instantaneous frequency it also needs
 a hypothesis that the positive derivative of $\theta(t)$ exists.

\vskip 1mm
\emph{In fact, no matter how the integral transform is defined, it is actually a uniform running-mean processing.
Now that the processing is done on the data, why not calculate the instantaneous frequency from the data
in a direct way? }
Historically speaking, there are only crude estimation methods for the frequency change.
Just as reviewed by Huang \emph{et al.} (2009a),
there is a fundamental ``zero-crossing method" which has been
used for a long time to compute the mean period (frequency) for narrow band data. Of
course, this approach is only meaningful for mono-component functions, where
the numbers of zero-crossings and extreme points must be equal in the data.
Huang \emph{et al.} had generalized the zero-crossing method
by improving the temporal resolution to a quarter wave period
with a running-mean approach.
This improvement yields a better estimation for the frequency, but
it is still incapable of reflecting the instantaneous changes.

\begin{figure}[!htbp]
\centerline{\hbox{\epsfig{figure=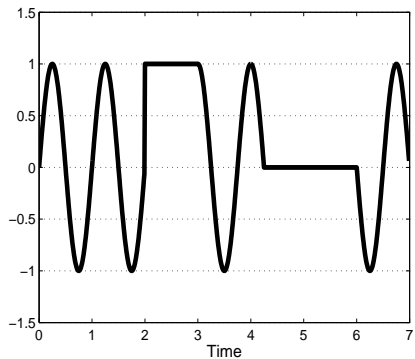,height=2in,width=2.5in,angle=0}}}
\caption{\footnotesize An example of periodic oscillation with intermittences.}
\end{figure}

\vskip 1mm
As an instantaneous frequency, it should be capable of reflecting the intermittent case
as in Fig.1, rather than excluding the adjacent-equal situation.
While the oscillation maintains unchanged at the extrema, the zero-crossing
or any other locations its instantaneous frequency should be all $0$.
Certainly, the frequency may lose its meaning on the junctures (belong to a null set).
It doesn't matter, after all, the data itself is not smooth.
\emph{Now that the period should be defined relative to a segment of time and
the frequency needs to be understood point by point, we can conciliate this
 conflict by an interpolating method.}
With this understanding, we have developed a ``direct interpolating" approach for
the calculation of the instantaneous frequency which will be illustrated in a latter section.

\subsection{About the Definition of IMF}

The EMD method defines an intrinsic mode function (IMF) with the following two
conditions [Huang \emph{et al.} (1998), Huang and Shen (2005)]:\\[1mm]
\textit{(1) In the whole data set, the number of extreme points and the number of zero-crossings
must either equal or differ at most by one.\\
(2) At any point, the mean value of the envelope defined by the local maxima and
the envelope defined by the local minima is zero.}

\vskip 1mm
The first condition can be understood as: the IMF's local maxima and minima points are septal
with no adjacent zero-crossings, all its maxima should be positive and all its minima should be negative.
Just as stated by Huang \emph{et al.} (1998), this request on oscillating manner
is for the rationality of defining an instantaneous frequency.
According to this restriction, the function given in Fig.1 is not an IMF.
But from the viewpoint of physics, the occurrence of this intermittence phenomenon is possible.
Therefore, it is reasonable for liberalizing the restriction and counting it as an IMF.

\vskip 1mm
The second condition requires that an IMF must have an envelope symmetry.
This restriction is for convenience of deducing a meaningful instantaneous frequency
by means of the Hilbert transform. Now that the Hilbert transform is abandoned
and the ``direct interpolating" approach is adopted, the restriction should be liberalized.
In fact, just as stated in our previous paper [Wang and Li (2012a)] the rigid
envelope-symmetric IMFs are out of reach for an actual data processing test.
Therefore, only if a decomposed component satisfies a certain error requirement
it can be seen as an IMF. In addition,
notice that the odd-even type of the extreme-point symmetry for ESMD\_II
is almost equivalent to the envelop symmetry for EMD and
the three-curve type for ESMD\_III is more general than the envelop symmetry,
this request can be also liberalized.

\vskip 1mm Based on the above analysis, the definition of IMF can be extended
with the following two conditions:\\[1mm]
\textit{(1) To count all the adjacent equal extreme points as one,
the local maxima and minima points should be septal,
all its maxima should be positive and all its minima should be negative.\\
(2) It should be almost envelop-symmetric or
  extreme-point symmetric in the generalized sense.}\\[1mm]
We note that the envelop symmetry and the odd-even type of the extreme-point symmetry are very good types
which yield IMFs with suitable amplitude and frequency modulations, and too low request on symmetry
would lead to difficulty to frequency and energy analysis.

\section{Decomposition Algorithm for ESMD Method}

We begin to introduce the ESMD method with the decomposition algorithm.
In the present paper, only one-dimensional data are considered.
 Certainly, there is a precondition for the processing that
the sampling rate of the observational instruments should be known.
It is a common sense that the local maxima and minima points are septal
with counting all the adjacent equal extreme points as one.
For convenience of programming, in case there are several adjacent equal extreme points,
we only choose the first one as a representative.
The program code is exploited on the Scilab platform and the algorithm is as follows:

\vskip 1mm\noindent\textbf{Step 1}: Find all the local extreme points (maxima points plus minima points) of the data $Y$ and
numerate them by $E_i$ with $1\leq i\leq n$.

\vskip 1mm
\noindent\textbf{Step 2}: Connect all the adjacent $E_i$ with line segments and
mark their midpoints by $F_i$ with $1\leq i\leq n-1$.

\vskip 1mm
\noindent\textbf{Step 3}: Add a left and a right boundary midpoints $F_0$ and $F_n$ through a certain approach.

\vskip 1mm
\noindent\textbf{Step 4}: Construct $p$ interpolating curves $L_1,\cdots, L_p$ ($p\geq 1$) with all these $n+1$ midpoints
and calculate their mean value by $L^*=(L_1+\cdots+L_p)/p$.

\vskip 1mm
\noindent\textbf{Step 5}: Repeat the above four steps on $Y-L^*$ until $|L^*|\leq \varepsilon$
($\varepsilon$ is a permitted error) or the sifting times attain a preset maximum number $K$. At this time, we get
the first mode $M_1$.

\vskip 1mm
\noindent\textbf{Step 6}: Repeat the above five steps on the residual $Y-M_1$ and get $M_2, M_3\cdots$ until
the last residual $R$ with no more than a certain number of extreme points.

\vskip 1mm
\noindent\textbf{Step 7}: Change the maximum number $K$ on a finite integer interval
$[K_{min}, K_{max}]$ and repeat the above six steps.
Then calculate the variance $\sigma^2$ of $Y-R$ and plot a figure with $\sigma/\sigma_0$ and $K$,
here $\sigma_0$ is the standard deviation of $Y$.

\vskip 1mm
\noindent\textbf{Step 8}: Find the number $K_0$ which accords with minimum  $\sigma/\sigma_0$
 on $[K_{min}, K_{max}]$. Then use this $K_0$
to repeat the previous six steps and output the whole modes. At this time, the last residual $R$ is actually an optimal
AGM curve.

\vskip 1mm There are several questions associated with this algorithm. In the following we explain them one by one.

\vskip 1mm
 According to the fourth step, we classify the ESMD into ESMD\_I, ESMD\_II, ESMD\_III, $\cdots$.
  ESMD\_I does the sifting process by using only $1$ curve interpolated
by all the midpoints; ESMD\_II does the sifting process by using $2$ curves interpolated
by the odd and even midpoints, respectively; ESMD\_III does the sifting process by using $3$ curves interpolated
by the midpoints numerated by $3k+1$, $3k+2$ and $3(k+1)$ ($k=0,1,\cdots$), respectively.
Certainly, we can also define other schemes with more interpolating curves according to this method.

\vskip 1mm In the fifth step, besides the permitted error $\varepsilon$, we can also adjust the maximum sifting times $K$.
 On the one hand, if $\varepsilon$ is the unique controlling parameter,
 it may leads to an endless loop to the decomposition; On the other hand, if $K$ is the unique controlling parameter,
 we know nothing about the symmetric properties of each mode.
 Perhaps a small number of sifting may lead to good symmetric to a mode. So the wise choice is to use them all.
 To obtain a series of relatively reliable modes, we can fix $\varepsilon$ to be a very small value and control the
 decomposing process by changing $K$ on a finite integer interval such that the last residual $R$ (AGM curve) is an optimal one.
 So Step $7$ and $8$ are very necessary.
 In fact, only when the fitting curve of the data is an optimal one, the remainder can be seen as
 an actual oscillation caused by a series of wave fluctuations.

\vskip 1mm Denote the original data and the AGM curve by $Y=\{y_i \}_{i=1}^N$
and $R=\{r_i \}_{i=1}^N$, respectively.
Commonly, relative to its total mean $\overline{Y}=\sum_{i=1}^N y_i/N$ the variance of the data is defined as
\begin{eqnarray}
\sigma_0^2=\frac{1}{N}\sum\limits_{i=1}^N (y_i-\overline{Y})^2.
\end{eqnarray}
Here we define the variance relative to the AGM by
\begin{eqnarray}
\sigma^2=\frac{1}{N}\sum\limits_{i=1}^N (y_i-r_i)^2.
\end{eqnarray}
In the applications, we usually choose $\varepsilon=0.001\sigma_0$ and
use the ratio $\nu=\sigma/\sigma_0$ to reflect the degree of optimization
for the AGM relative to the common total mean.

\vskip 1mm In addition, the third step is associated with a boundary processing which is a ``benevolent see benevolence" problem.
In our program codes we have developed the linear interpolation method given by Wu and Huang (2009) and revised
the interpolation styles for the too steep boundary case [see \emph{Appendix A}].
This revision can make the boundary much more stable, even for the tests with $100,000$
sifting times [Wang and Li (2012a)].
In the following we test the decomposition effectiveness
 according to ESMD\_I, ESMD\_II and ESMD\_III
and our attention is mainly focused on the second one.

\section{Performance of ESMD\_I}

ESMD\_I does the sifting process by using only $1$ curve interpolated by all the midpoints of the line segments between the local maxima and minima points.
Though this case has been discussed by Huang \emph{et al.}(1998), it is worth reemphasizing
from the viewpoint of ESMD. We test it by a simple example as follows.
\vskip 1mm
\noindent\textbf{Example 1}: $Y(t)=e^{-0.1t}\sin{(\pi t/2+\pi/3)}$ with $0\leq t\leq 20$.
\vskip 1mm
This is a weighted-periodic function with a fixed frequency and varying amplitude.
A data may maintain its oscillating frequency and increase (or decrease) its amplitude
with gaining (or losing) energy. So it is a very natural thing to meet
 the weighted-periodic function in data processing. A good sifting scheme is anticipated
yielding a unique IMF with a small decomposition error.

\begin{figure}[!htbp]
\begin{minipage}[t]{0.45\linewidth}
\centerline{\hbox{\epsfig{figure=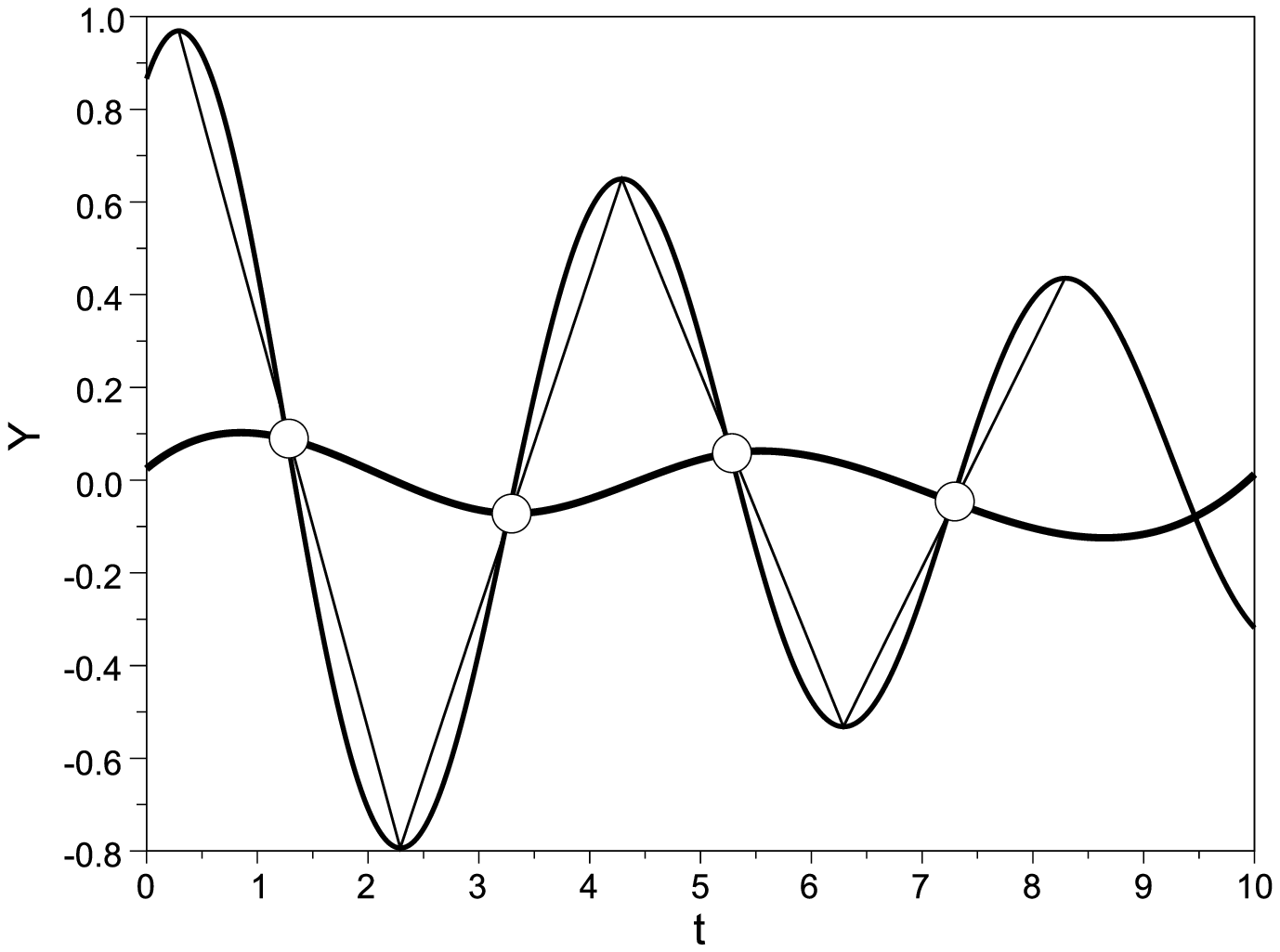,height=2.1in,width=2.5in,angle=0}}}
\caption{\footnotesize All the midpoints (circle points) of the line segments between the local maxima and minima points and their unique
interpolating curve (bulk inner curve).}
\end{minipage}
\hspace{3ex}
\begin{minipage}[t]{0.45\linewidth}
\centerline{\hbox{\epsfig{figure=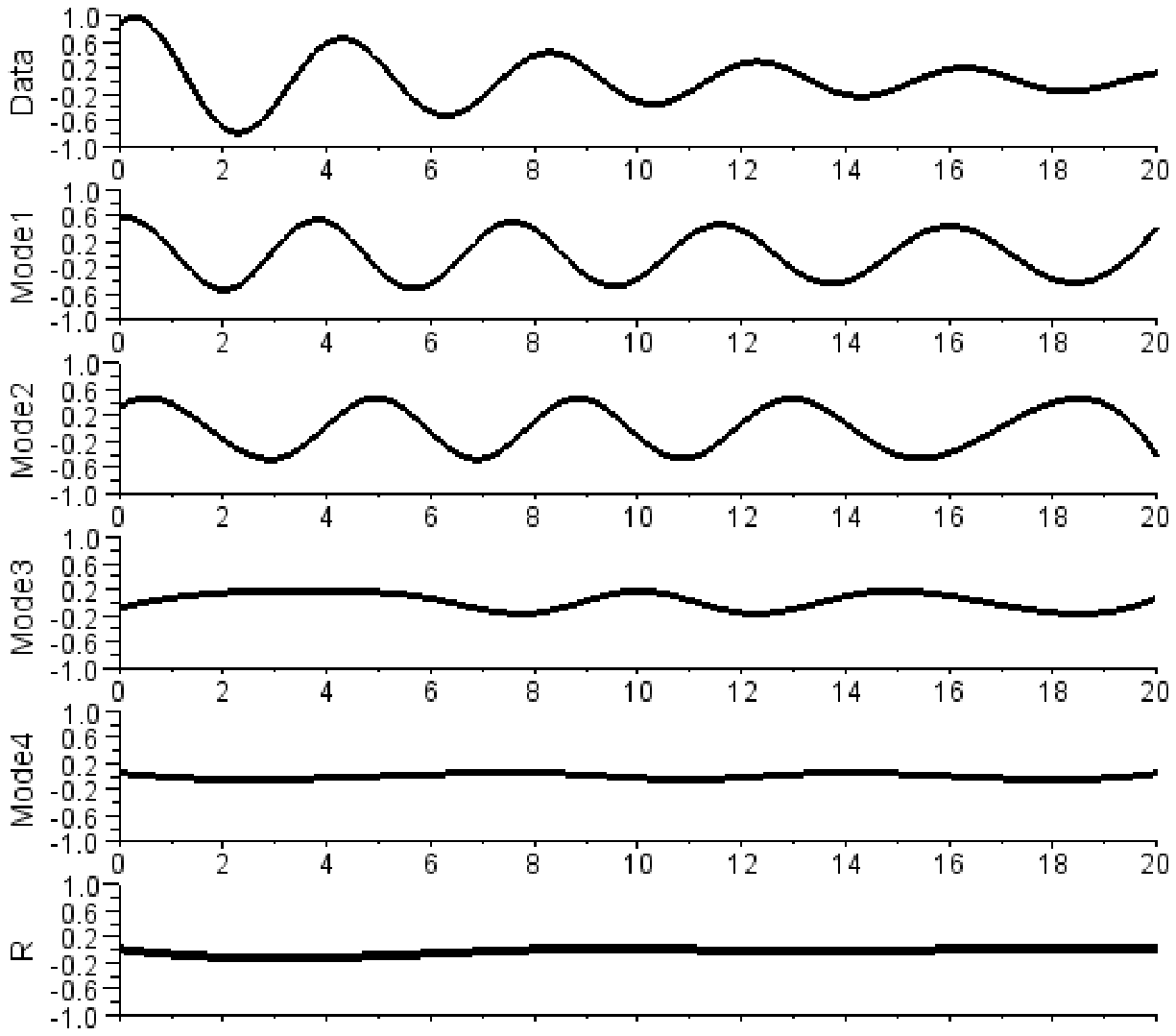,height=2.2in,width=2.5in,angle=0}}}
\caption{\footnotesize The decomposition result for the weighted-periodic function given by ESMD\_I with 20 sifting times, here
the horizontal axis stands for the time (second).}
\end{minipage}
\end{figure}

\vskip 1mm The detailed constructing process of the interpolating curve is shown in Fig.2.
\emph{If there is no variance-ratio investigations, we know nothing about the effect of sifting times.}
To try the decomposition with $20$ times sifting, it yields a result in Fig.3, which includes $4$ modes and a residual R.
There is a common feature for these modes that all their amplitudes almost maintain unchanged.
In fact, it is due to the scheme of the rigid extreme-point symmetry adopted by ESMD\_I.
Through a simple geometric proof we see a curve
with an equal amplitude in the form $A\sin{\theta(t)}$ ($\theta(t)$ is an increasing function) must be an extreme-point symmetric one.
Vice versa, it is also true.
Particularly, the Mode1 is not only extreme-point symmetric but also periodic. In fact, it is an approximation
of the function $0.6\sin{(\pi t/2+\pi/3)}$ which carries the most periodicity of the original data.
This try shows that $20$ times sifting accords with an enough-symmetry and slow-efficiency case.
Hence, a lower times with lose symmetry request may be better.
but there is no effective criteria for it under the EMD circumstance.
One progress of ESMD is on the adoption of variance ratio.

\vskip 1mm Firstly, we plot the distribution figure of the variance ratio $\nu=\sigma/\sigma_0$ along the sifting times (see Fig.4)
 and find out the optimal sifting time $3$ which accords with the minimum value $\nu=99.8\%$;
 Secondly, we output the corresponding decomposition result with $3$ sifting times.
It follows from Fig.5 that the result is better than that in Fig.3 with $20$ times sifting,
where Mode1 can be seen as an approximation
of the original data and the others can be seen as the decomposition error. Certainly, this decomposition
is not perfect either, after all, the error amplitude achieves $0.3$.
\begin{figure}[!htbp]
\begin{minipage}[t]{0.45\linewidth}
\centerline{\hbox{\epsfig{figure=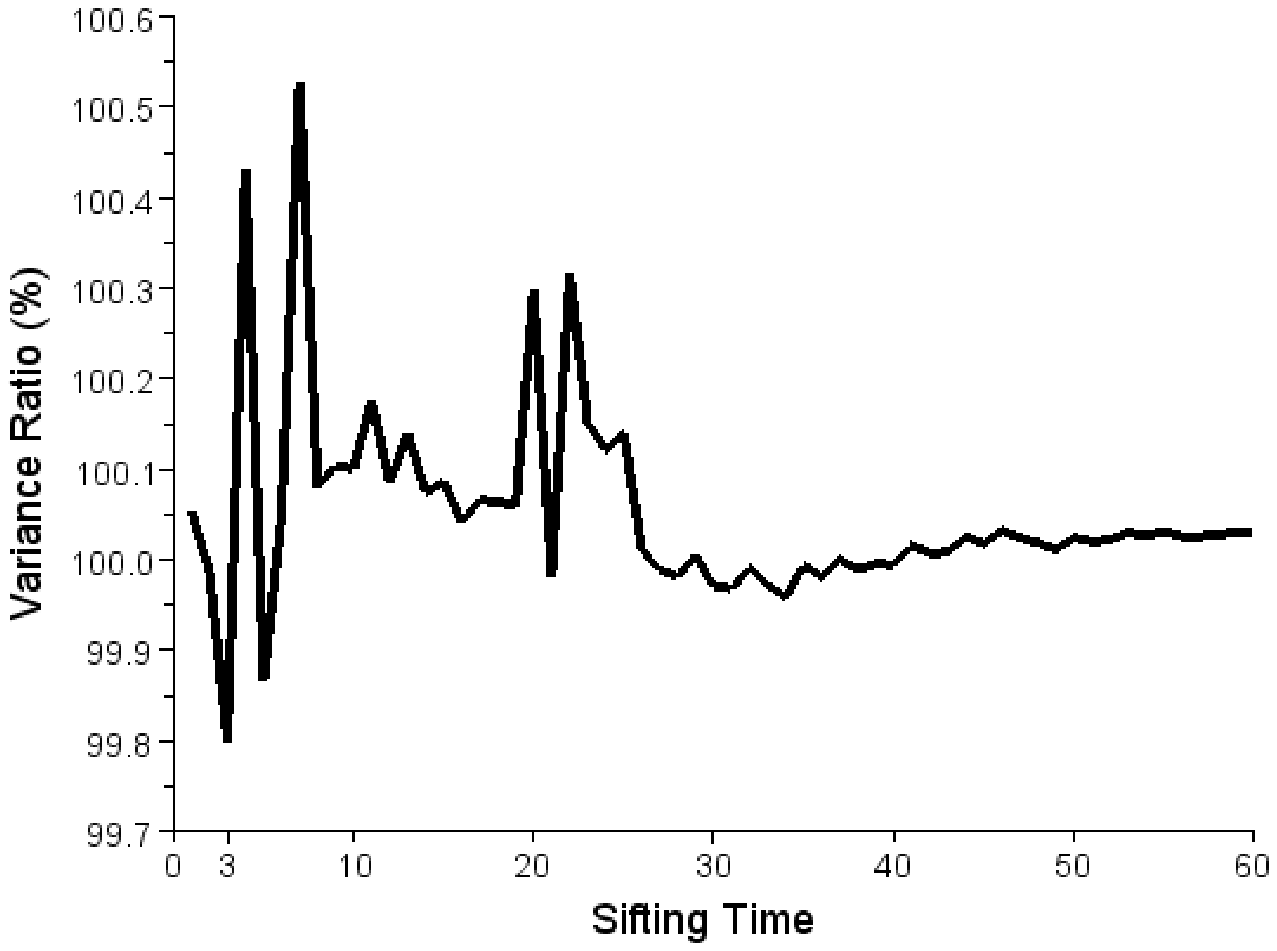,height=2in,width=2.5in,angle=0}}}
\caption{\footnotesize The distribution of variance ratio $\nu=\sigma/\sigma_0$ along
the sifting times for the weighted-periodic function.}
\end{minipage}
\hspace{4ex}
\begin{minipage}[t]{0.45\linewidth}
\centerline{\hbox{\epsfig{figure=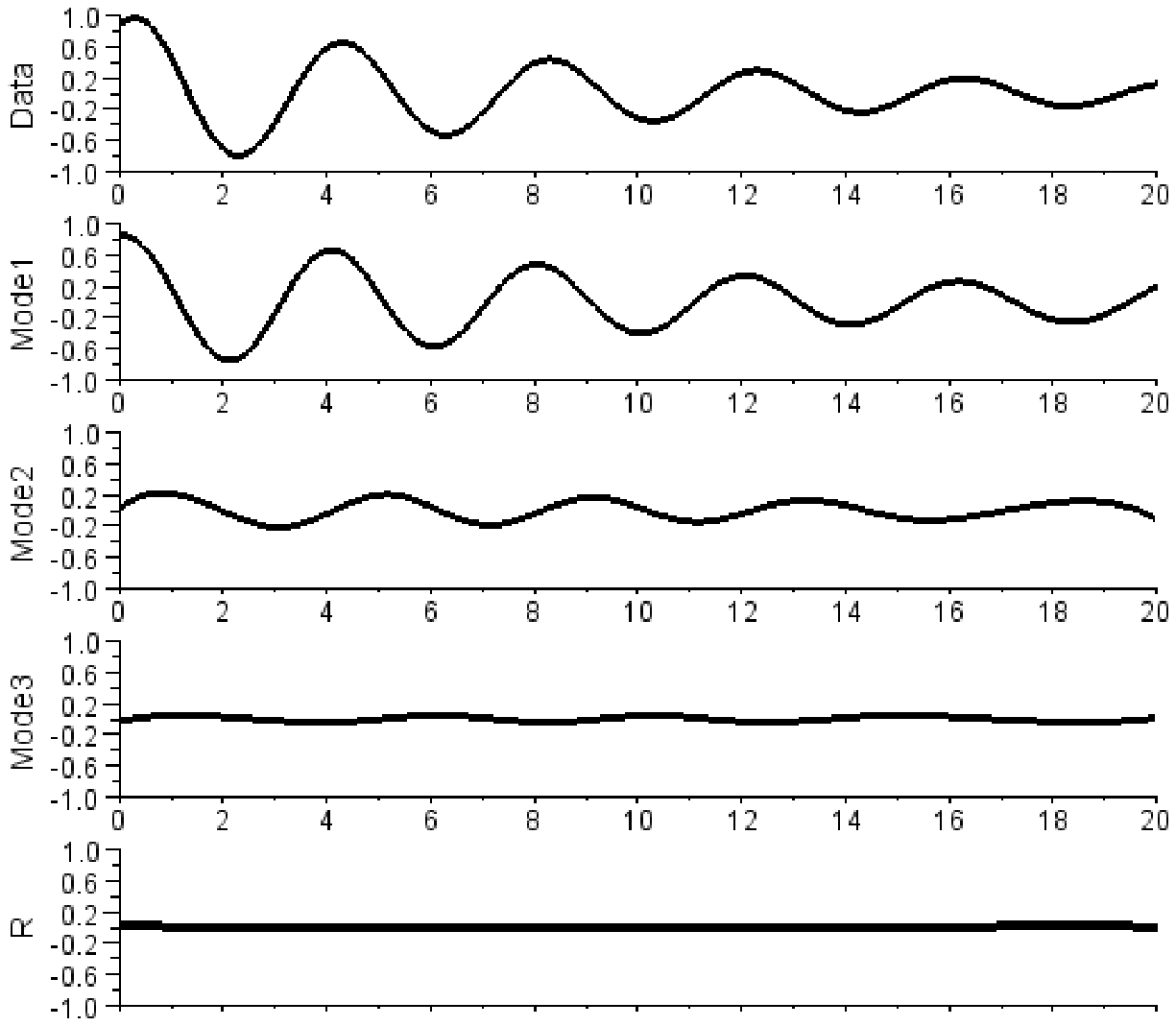,height=2.2in,width=2.5in,angle=0}}}
\caption{\footnotesize The decomposition result for the weighted-periodic function given by ESMD\_I with $3$ sifting times, here
the horizontal axis stands for the time (second).}
\end{minipage}
\end{figure}

\vskip 1mm
\noindent\textbf{Example 2}: A segment of wind data observed at sea with $20$Hz sampling rate.
\vskip 1mm

It follows from Fig.6 that $18$ is the optimal sifting times in the interval $[1, 30]$.
The corresponding decomposition in Fig.7 yields $12$ IMFs together with a residual R
with $40\%$ variance ratio. It means the AGM is the best fitting curve of the wind data.
 At this time, the IMFs still have an amplitude-modulation phenomenon.
If the sifting times is added, more and more equal-amplitude IMFs may appear.
In consideration of physical meaning stated by Huang \emph{et al.} (2003),
we do not expect too many equal-amplitude modes. So the lower sifting times with
imperfect decomposition is preferred for ESMD\_I.
\begin{figure}[!htbp]
\begin{minipage}[t]{0.45\linewidth}
\centerline{\hbox{\epsfig{figure=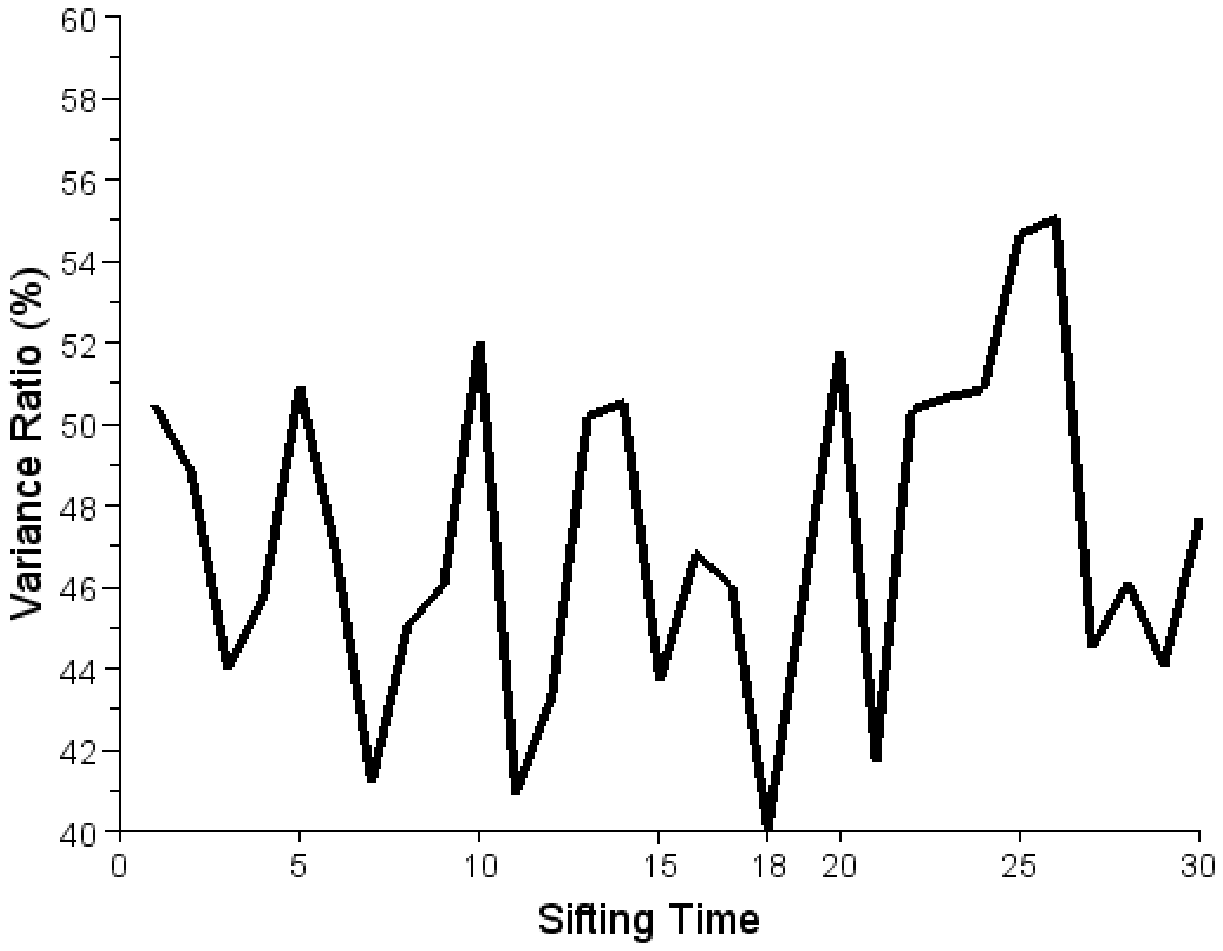,height=2.0in,width=2.3in,angle=0}}}
\caption{\footnotesize The distribution of variance ratio $\nu=\sigma/\sigma_0$ along the sifting times for the wind data.}
\end{minipage}
\hspace{4ex}
\begin{minipage}[t]{0.45\linewidth}
\centerline{\hbox{\epsfig{figure=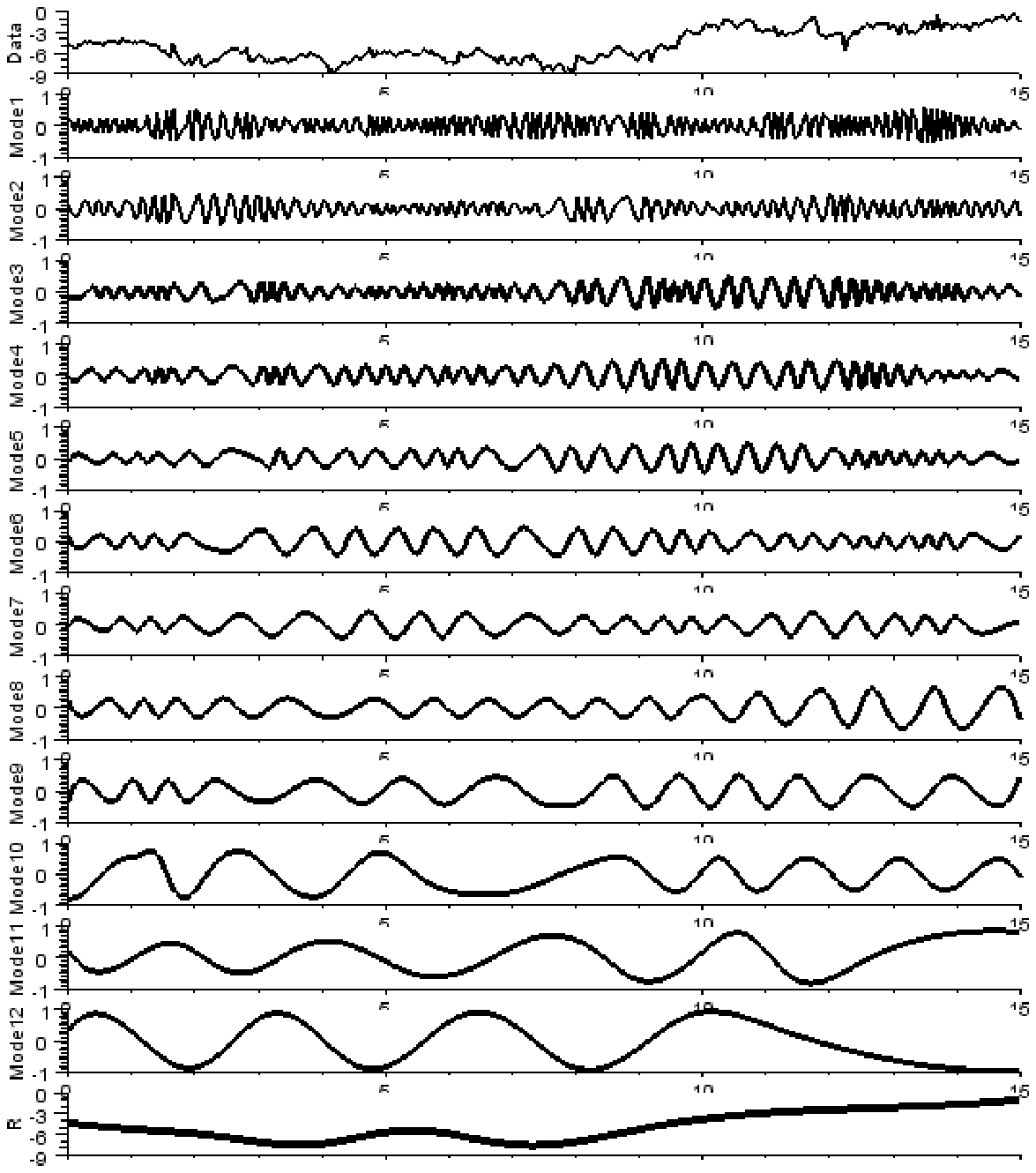,height=3.0in,width=2.5in,angle=0}}}
\caption{\footnotesize The decomposition result for the wind data given by ESMD\_I with $18$ sifting times, here
the horizontal axis stands for the time (second).}
\end{minipage}
\end{figure}

\vskip 1mm
In all, due to the adoption of a scheme with the rigid extreme-point symmetry,
the decomposition efficiency of ESMD\_I is not high.
Besides the rigorous request of symmetry, there is also another aspect. It follows
from Fig.2 that when all of the midpoints are used for interpolating, the generated curve may contain
almost the same number of extreme points
as the original data which may subsequently enter into the second mode. This also leads to low efficiency to the decomposition.
Though ESMD\_I has this defect, its AGM curve may be very good.
Comparatively, the EMD method has a relatively high decomposing efficiency. One reason is that,
the request of envelope symmetry is lower than that of the rigid extreme-point symmetry; Another reason is that,
the upper and lower envelopes are interpolated by almost half number of the data's extreme points and
their mean curve discounts the number by almost a half. In view of these facts, it is very natural to
do the decomposition by using $2$ interpolating curves.

\section{Performance of ESMD\_II}

ESMD\_II does the sifting process by using $2$ curves interpolated
by the odd and even midpoints, respectively. The detailed constructing process of the curves is shown in Fig.8.
For this case, the decomposition of \emph{Example 1}
is trivial since the original curve itself is a permitted mode.
In the following we test ESMD\_II by three examples
and analyze its characteristics in three aspects.
\begin{figure}[!htbp]
\centering
\includegraphics[width=2.2in]{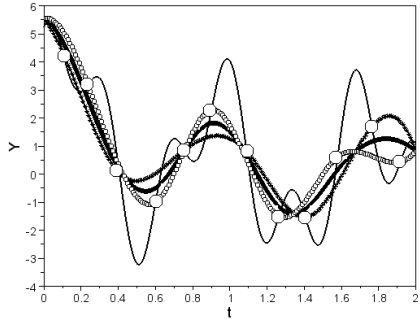}
 \caption{All the midpoints (big circle points) of the line segments between the local maxima and minima points
of the data (thin solid curve) and the mean curve (thick solid curve) of their odd and even
interpolating curves. }
\end{figure}

\subsection{Decomposition Tests}
\vskip 1mm
\noindent\textbf{Example 3}: $Y(t)=-\sin(8\pi t)+1.5e^{-0.2 t}\sin{(1.9 \pi t+\pi/20)}+(t-2)^2$, $0\leq t\leq 4$.
\vskip 1mm
This data is composed of one periodic function, one weighted-periodic function and one parabola.
In the following we do the decomposition with ESMD\_II.
  From Fig.9 and 10 we see the decomposition is perfect.
  Mode1 accords with the periodic function; Mode2 accords with the weighted-periodic function;
 The residual R accords with the parabola.
\begin{figure}[!htbp]
\begin{minipage}[t]{0.45\linewidth}
\centerline{\hbox{\epsfig{figure=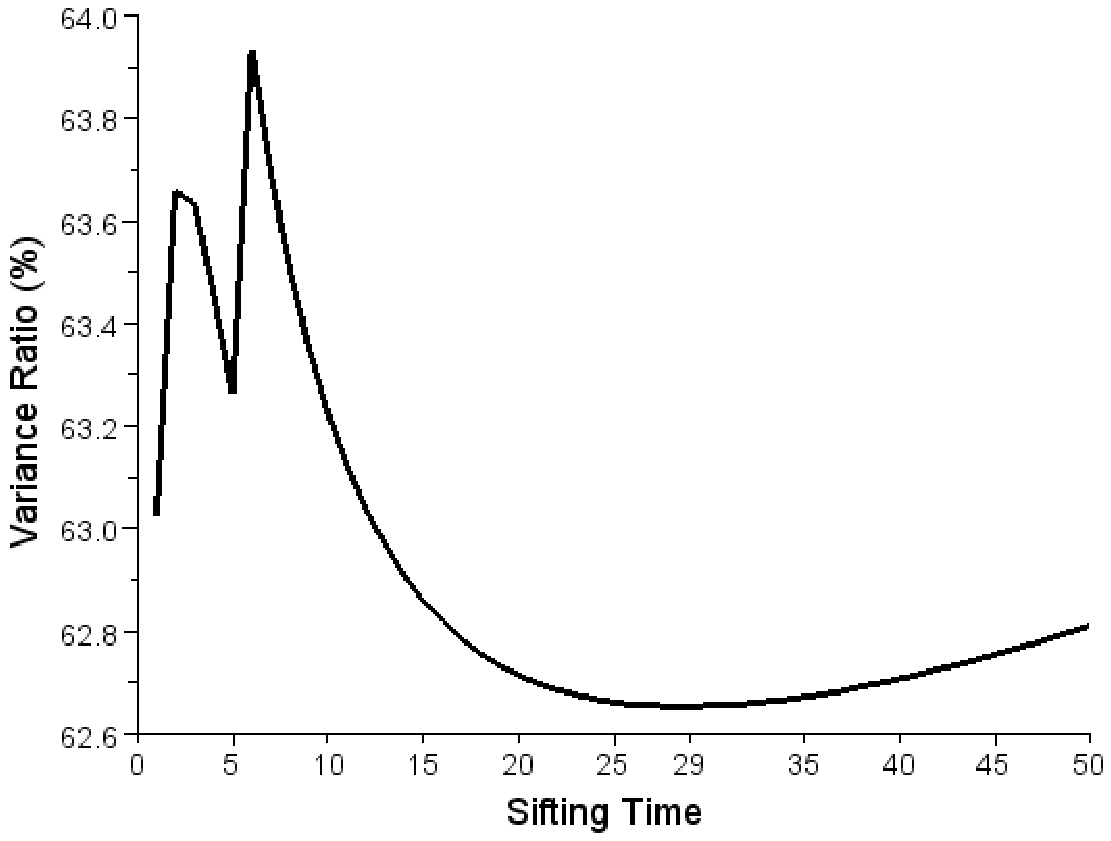,height=2in,width=2.5in,angle=0}}}
\caption{\footnotesize The distribution of variance ratio $\nu=\sigma/\sigma_0$ along the sifting times for the composed data.}
\end{minipage}
\hspace{3ex}
\begin{minipage}[t]{0.45\linewidth}
\centerline{\hbox{\epsfig{figure=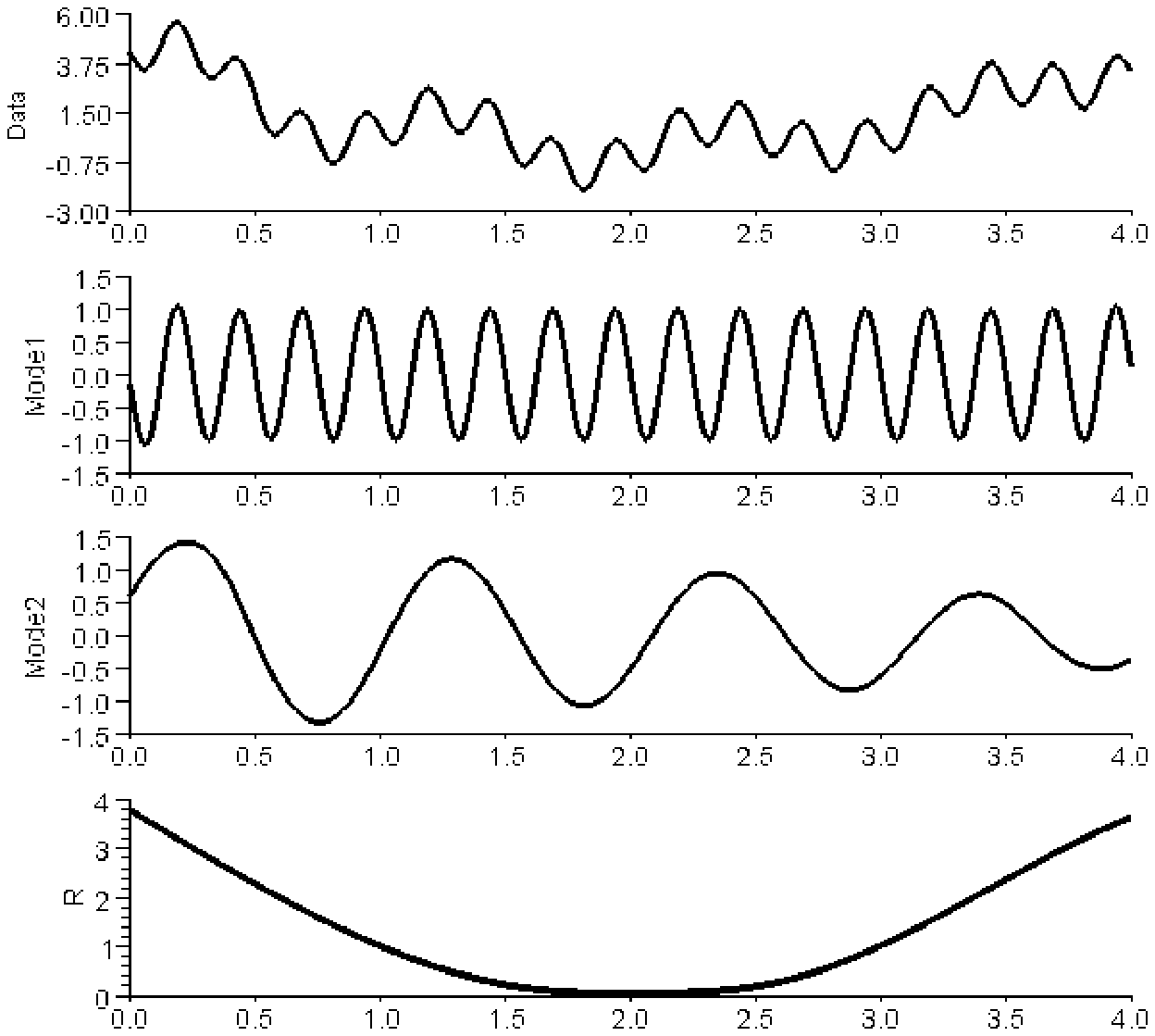,height=2.1in,width=2.5in,angle=0}}}
\caption{\footnotesize The decomposition result for the composed data given by ESMD\_II with $29$ sifting times, here
the horizontal axis stands for the time (second). }
\end{minipage}
\end{figure}

\vskip 1mm
In the following we re-decompose Example 2 with ESMD\_II.
From Fig.11 we see the variance ratio $\nu=\sigma/\sigma_0$ attains its
minimum value at $30$, which can be seen as an optimal sifting times
in the whole interval $[1,100]$. Fig.12 shows the corresponding
decomposition which is more distinct than that in Fig.7 given by ESMD\_I.
 The decomposed IMFs accord with the components of the wind turbulence with average periods
$3.8$s, $1.5$s, $0.6$s,$\cdots$. In addition, the last residual R is an
optimal AGM curve, which reflects the fundamental
evolutionary trend of the wind speed very well (see Fig.13).
Certainly, there is a necessity for us to compare with EMD method.
So the test is also done on the code eemd.m (downloaded from http://rcada.ncu.edu.tw/class2009.htm)
for the non-noise case. By the way, to be more objective, the default $10$ times sifting
given by Wu and Huang (2009) is substituted by $30$ here.
From Fig.12 and Fig.14 we see the difference is very clear.
This difference probably lies in the sifting scheme, the boundary processing
and the programming. Certainly, it is difficult for us to judge which IMF set is more reasonable.
But it follows from the residual comparison in Fig.13 we see
the global mean curve given by ESMD\_II is better than the monotone form given by the EMD method,
after all, this one is found by an optimizing approach. It also indicates that the AGM curve
is almost equivalent to the sum of Mode5, Mode6, Mode7 and the last trend function of EMD.
Generally speaking, the global mean curve is a component with maximum magnitude,
its deviation would lead to large distortion to the oscillating IMFs.
So in this sense, ESMD\_II is preferable.

\begin{figure}[!htbp]
\begin{minipage}[t]{0.45\linewidth}
\centerline{\hbox{\epsfig{figure=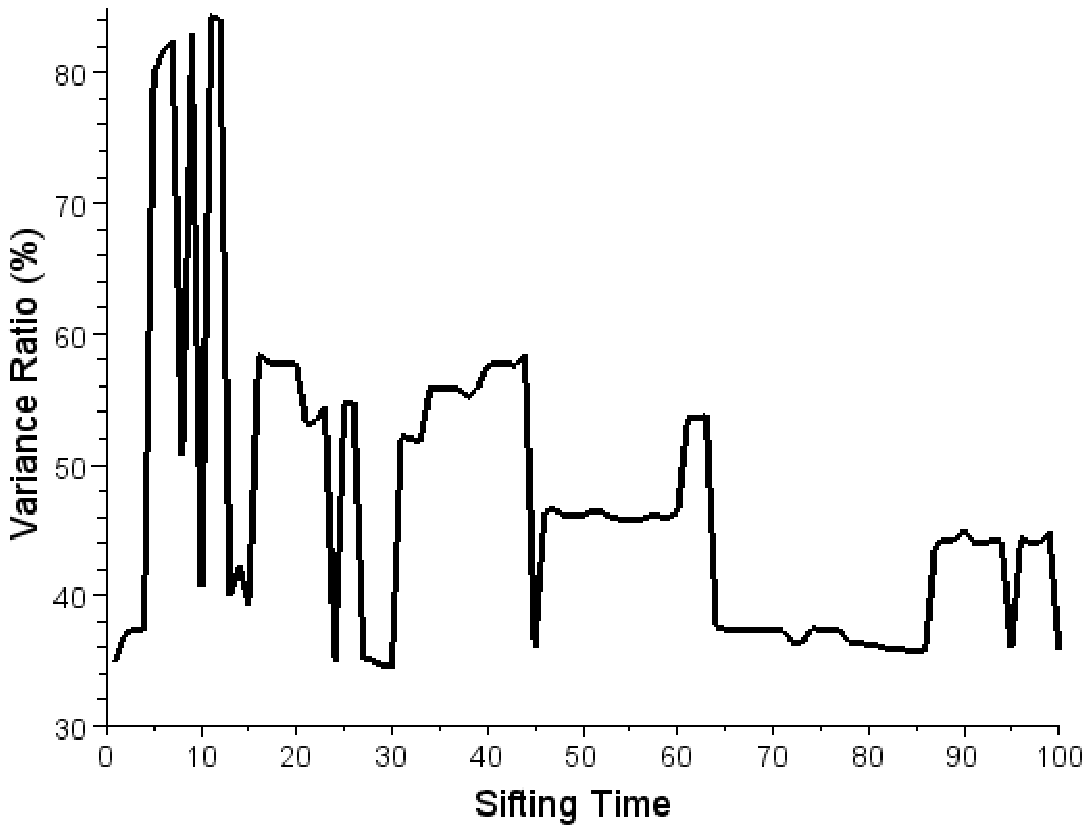,height=2in,width=2.4in,angle=0}}}
\caption{\footnotesize The distribution of the variance ratio $\nu=\sigma/\sigma_0$ along the sifting times for the wind data. }
\end{minipage}
\hspace{3ex}
\begin{minipage}[t]{0.45\linewidth}
\centerline{\hbox{\epsfig{figure=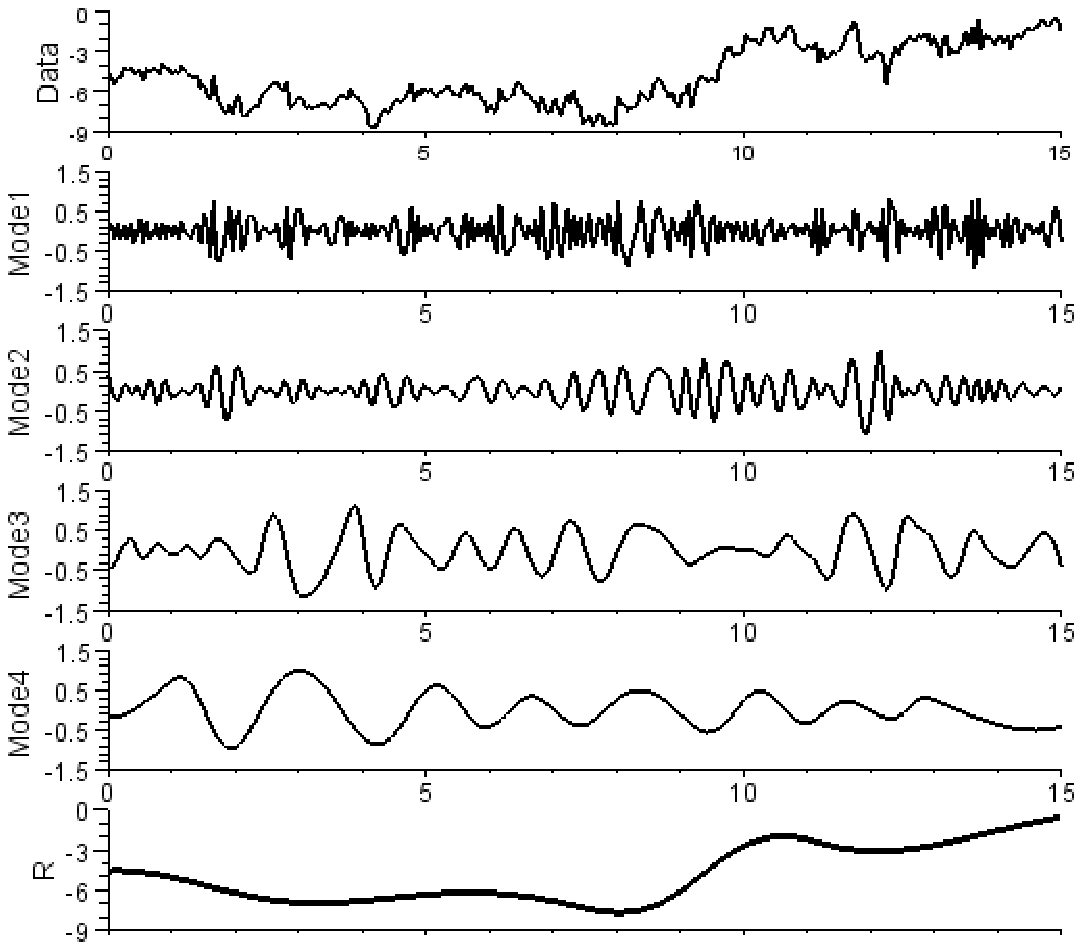,height=2.2in,width=2.4in,angle=0}}}
\caption{\footnotesize The decomposition result for the wind data given by ESMD\_II with $30$ sifting times, here
the horizontal axis stands for the time (second). }
\end{minipage}
\end{figure}
\begin{figure}[!htbp]
\begin{minipage}[t]{0.45\linewidth}
\centerline{\hbox{\epsfig{figure=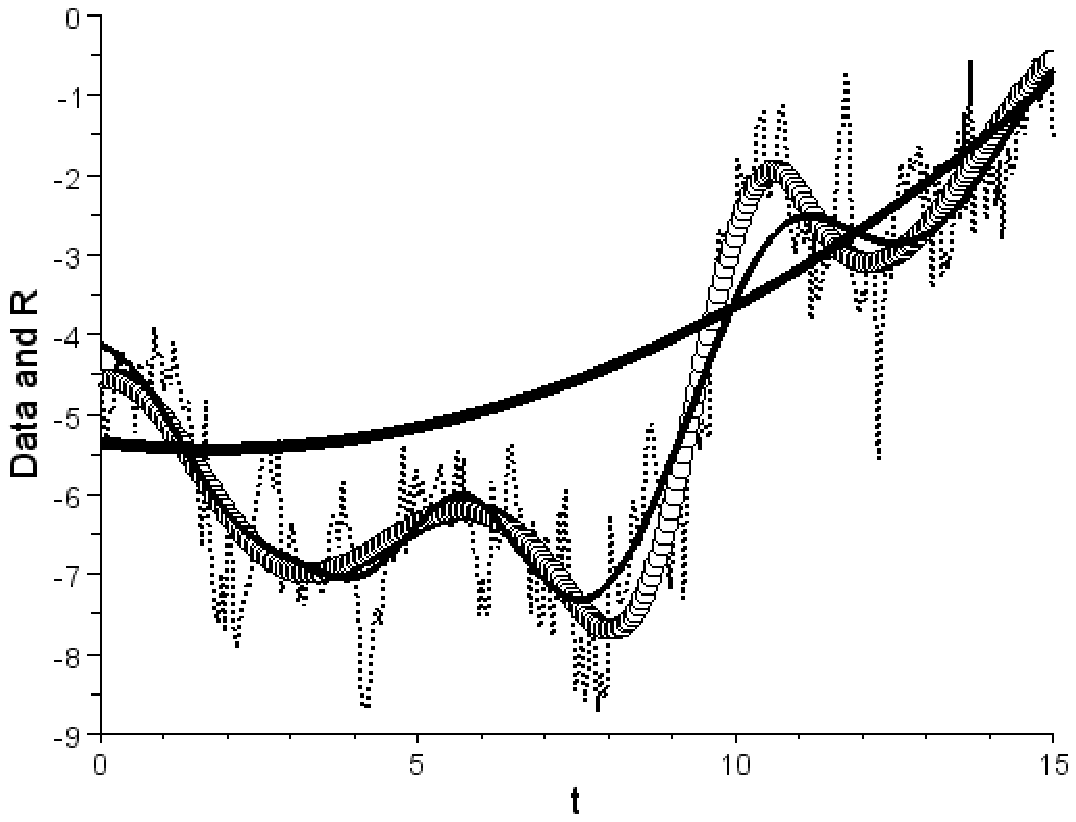,height=2in,width=2.4in,angle=0}}}
\caption{\footnotesize Comparison between the AGM curve of ESMD\_II (the curve with small circles),
the trend function (thick solid curve) as well as the sum of R and Mode5-7 of EMD (thin solid curve),
 where the dotted curve stands for the wind data. }
\end{minipage}
\hspace{3ex}
\begin{minipage}[t]{0.45\linewidth}
\centerline{\hbox{\epsfig{figure=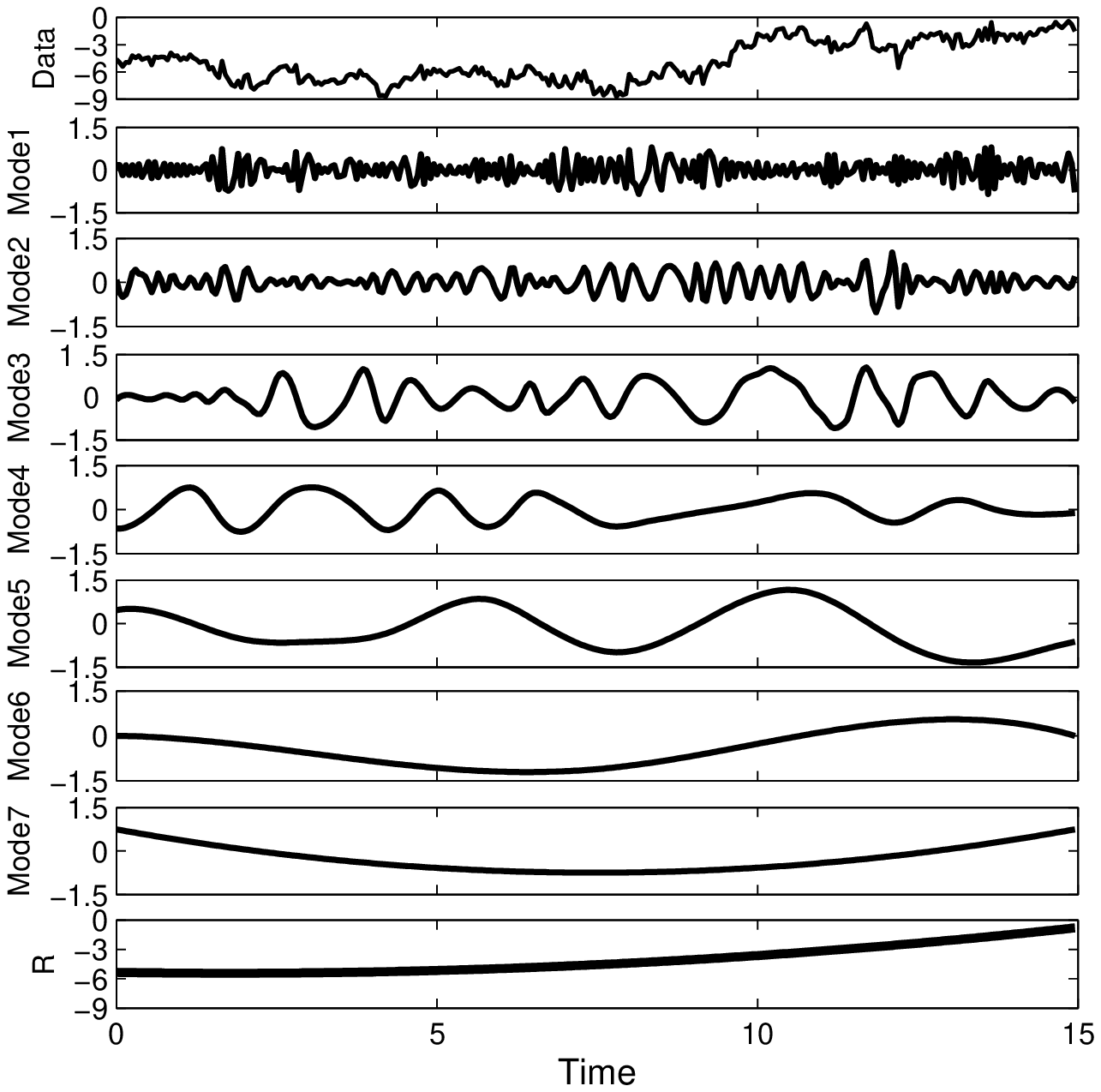,height=2.7in,width=2.5in,angle=0}}}
\caption{\footnotesize The decomposition result for the wind data given by EMD method with $30$ sifting times. }
\end{minipage}
\end{figure}

\noindent\textbf{Example 4}: The day-averaged air temperature data from May 10, 2008 to Nov. 3, 2011 downloaded from
the National Climatic Data Center of America.
\vskip 1mm
From Fig.15 we see the variance ratio has some septal stable intervals, such as $[20, 29]$, $[36, 40]$, $[43, 48]$ and
$[72, 76]$. On all these stable intervals, the variance ratio is almost identical, which leads
to almost the same result to the decomposition.
At this time the optimal sifting time is $29$,
even when the whole interval is prolonged to $[1, 200]$. The decomposition in Fig.16 shows a very good seasonal evolutionary trend for the air temperature.
The decomposed IMFs accord with the components with average periods
$66$ day, $35$ day, $17$ day,$\cdots$, which can be understood as bimonthly, monthly, semimonthly,$\cdots$ temperature oscillations. We refer the readers to Huang \emph{et al.} (2009b) and Bao \emph{et al.} (2011) for the analyzing
approach. Particularly, we can judge the period and time of the temperature anomaly from the amplitude variation. For this case, the amplitude variation of Mode5 is small, yet that of Mode4 is very big. These indicate that the temperature anomaly mainly occurs on time-scale of 35 days in Jan.-Mar., 2009.
\begin{figure}[!htbp]
\begin{minipage}[t]{0.45\linewidth}
\centerline{\hbox{\epsfig{figure=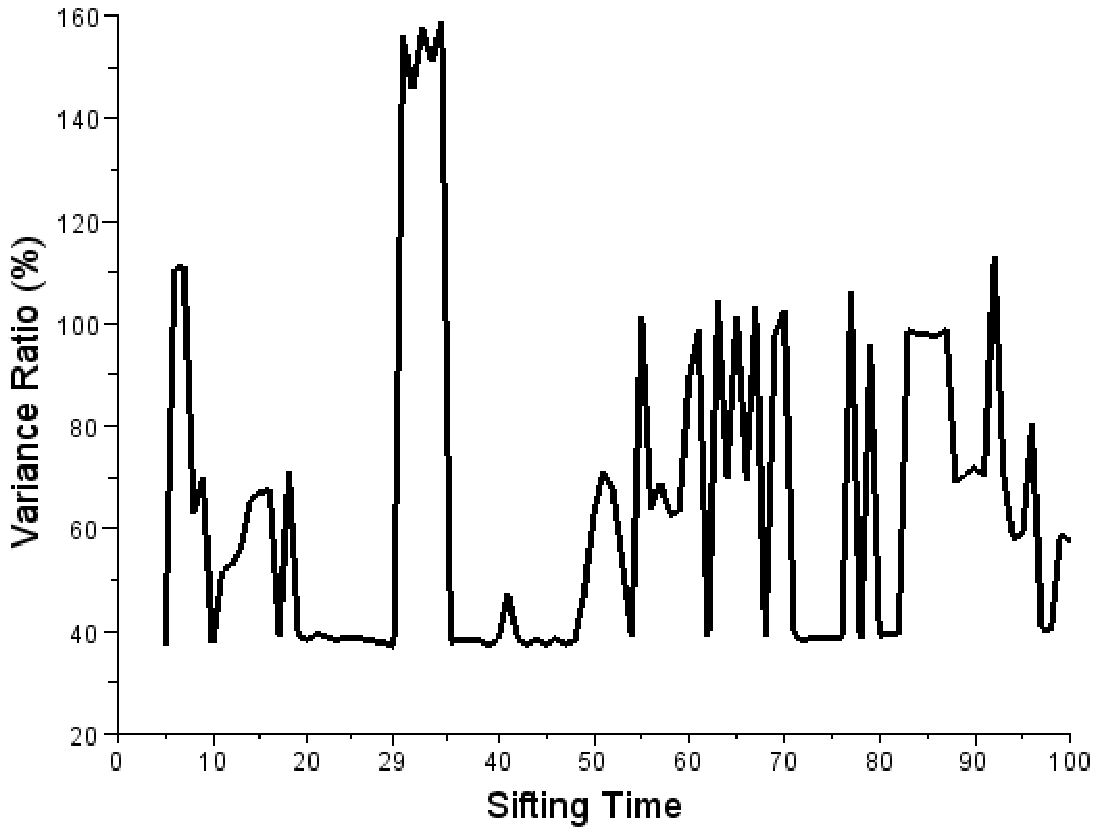,height=2in,width=2.5in,angle=0}}}
\caption{\footnotesize The distribution of variance ratio $\nu=\sigma/\sigma_0$ along the sifting times for the temperature data. }
\end{minipage}
\hspace{4ex}
\begin{minipage}[t]{0.45\linewidth}
\centerline{\hbox{\epsfig{figure=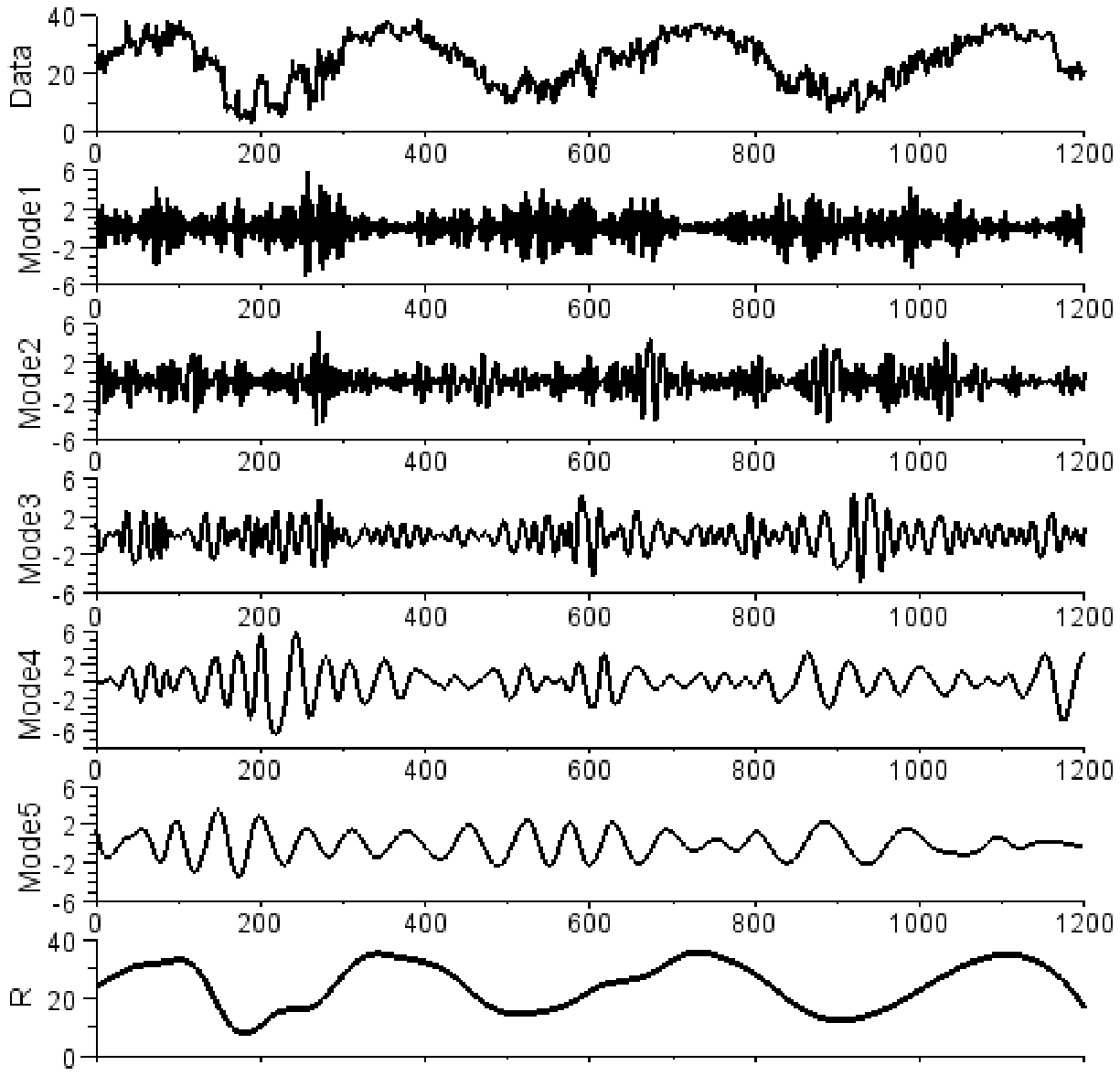,height=2.4in,width=2.5in,angle=0}}}
\caption{\footnotesize The decomposition result for the temperature data given by ESMD\_II with $29$ sifting times, here
the horizontal axis stands for the time (day).}
\end{minipage}
\end{figure}

\subsection{Symmetry Characteristic of the Modes}

 When the decomposed mode is a periodic function, such as Mode1 in Fig.10, its
symmetry characteristic is very obvious. For this case, bother
the odd and even interpolating curves and their mean value
all approximate the zero line. This is a real extreme-point symmetry.
But if the decomposed mode is a weighted-periodic function, such as Mode2 in Fig.10,
its symmetry is an extreme-point symmetry
of odd-even type, which only requires the symmetry between the odd and even interpolating curves.
In fact, this phenomenon is very universal. For an actual data component, not only
its amplitude but also its frequency changes along the time, such as Mode1 in Fig.12.
It follows from Fig.17 that this case is almost equivalent to the envelope symmetry.
However, since the magnitudes of the odd and even interpolating curves are smaller than that of the upper and lower envelopes,
the convergence speed of their mean curve to the zero line may be quicker.
So the ESMD\_II method may need less sifting times than that of the EMD method to reach a relatively stable state.

\subsection{Effect of Sifting Times to the Decomposition}

 According to the ESMD algorithm Step 4 and 5, adding the sifting times may make the modes
more and more symmetric until $|L^*|\leq \varepsilon$.
Moreover, since the permitted error $\varepsilon$ is preestablished (such as $\varepsilon=0.001\sigma_0$),
more times of sifting may imply more symmetric modes.
So we can anticipate a finite times of sifting such that all the modes satisfy this permitted error.
This case accords with a relatively stable variance ratio $\nu=\sigma/\sigma_0$. For Fig.9 the stable interval
is approximately $[25, 34]$; For Fig.11 and Fig.15 there are several septal stable intervals.
In the symmetry requirement we prefer to choose the optimal sifting times in the stable intervals,
though there may be some lower sifting times which accord with lower $\nu$.
By the way, it is not the case that more times of sifting leads to a better decomposition. On the one hand, as shown in Fig.9,
the additional sifting may lead to additional error to the decomposition; On the other hand, as shown in Fig.15, the decomposition
may be not uniform convergent about the sifting times.
Certainly, there is also another case that the stable interval does not appear for several hundred times of sifting.
This is possibly caused by a too small value of $\varepsilon$. At this time, we suggest replacing the value of $\varepsilon$
by a bigger one and redoing it. In addition, we note that in the stable interval the decomposition is insensitive to
the sifting times and, on the contrary, in the unstable interval it differs much. Especially, when the variance ratio $\nu>100\%$,
the corresponding decomposition may be very bad.

\subsection{Effect of Least Extreme-Point Number to the Decomposition }

 To stop the decomposition it requires the least number of extreme points.
This number, denoted by $m_R$, may influence the shape of R.
In order to satisfy the requirement of the linear interpolation on the boundaries, it requires $m_{R}\geq 4$.
The default one is $4$.
If the so-called optimal R for the default case has large difference from the data
(reflected by a very high $\nu$), we can increase $m_{R}$ and redo it
to get a better one. Fig.9 and 10 are the default decompositions; Fig.11 and 12 are the
 decompositions with $m_{R}=6$; Fig.15 and 16 are the
 decompositions with $m_{R}=8$.
 In Fig.16 if $m_{R}$ is increased to $20$ the AGM curve may become
 the combination of Mode5 and R which should be avoided, though the variance ratio is much lowered down at this time.
\begin{figure}[!htbp]
\begin{minipage}[t]{0.45\linewidth}
\centerline{\hbox{\epsfig{figure=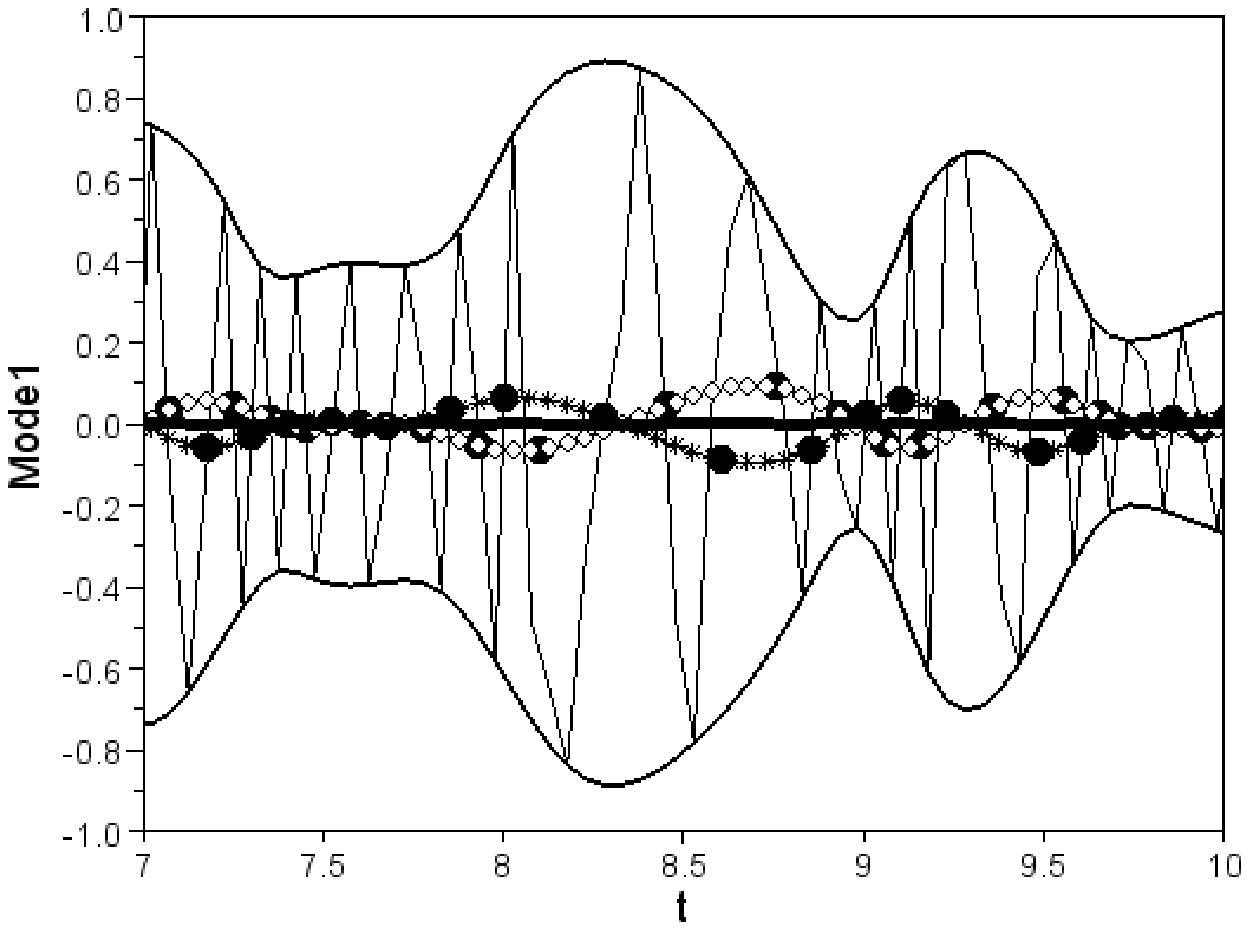,height=2in,width=2.5in,angle=0}}}
\caption{\footnotesize The odd-even type of extreme-point symmetry for Mode$1$ of Fig.12
(symmetry between the inner curves with $\ast$ and $\diamond$). Here two outer envelop curves are also shown.}
\end{minipage}
\hspace{4ex}
\begin{minipage}[t]{0.45\linewidth}
\centerline{\hbox{\epsfig{figure=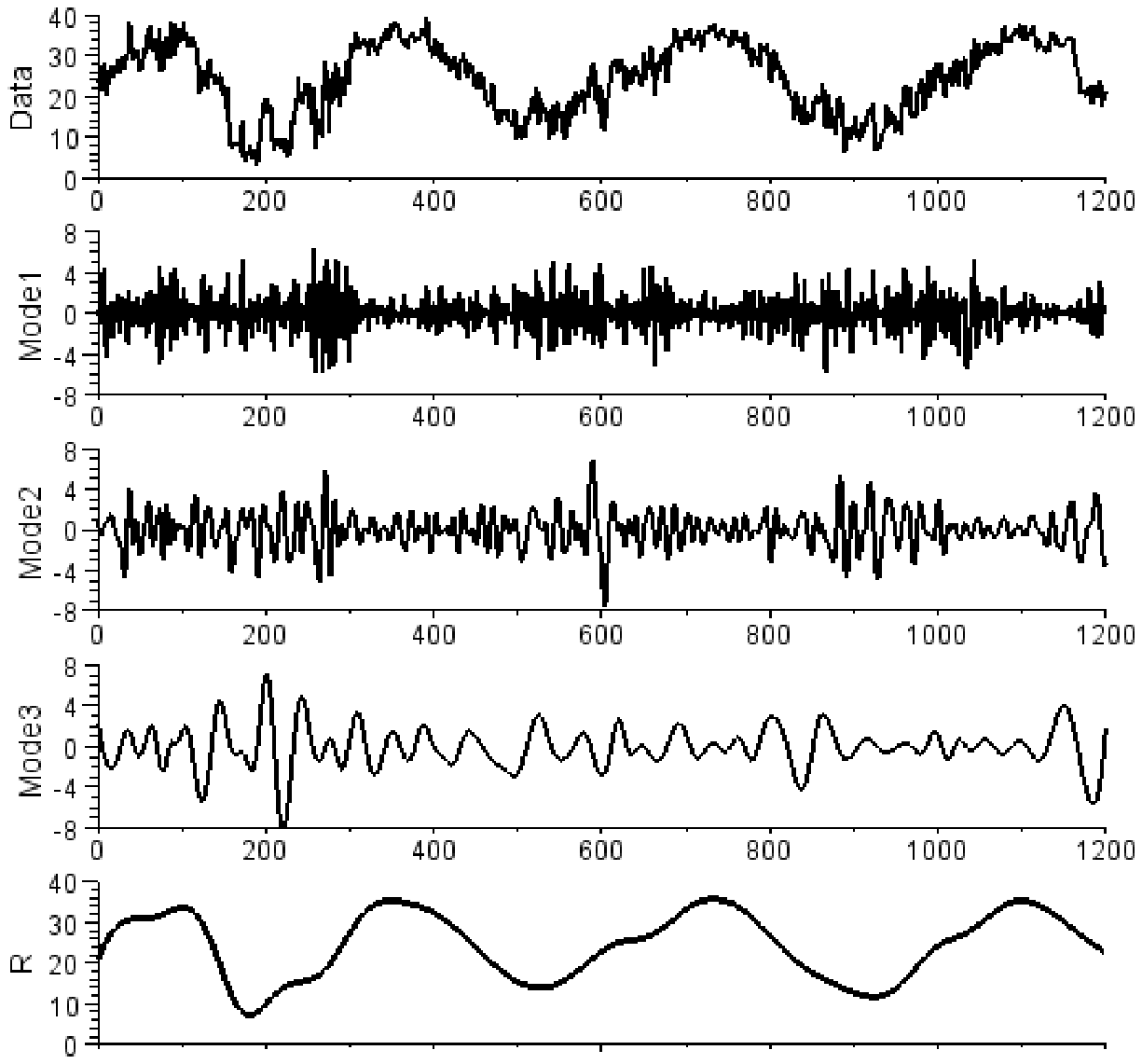,height=2.3in,width=2.5in,angle=0}}}
\caption{\footnotesize The decomposition result for the temperature data given by ESMD\_III with $11$ sifting times, here
the horizontal axis stands for the time (day).}
\end{minipage}
\end{figure}

\section{Performance of ESMD\_III}

 ESMD\_III does the sifting process by using $3$ curves $L_1, L_2$ and $L_3$ interpolated
by the midpoints numerated by $3k+1$, $3k+2$ and $3(k+1)$ ($k=0,1,\cdots$), respectively.
For this case the mean curve is defined as $L^*=(L_1+L_2+L_3)/3$. To make a mode symmetric it requires
$|L^*|\leq \varepsilon$. This kind of symmetry is an extreme-point symmetry
in the more extensive meaning, which only requires the symmetry between $L_1+L_2$ and $L_3$.
Particularly, if the component is a periodic function it becomes a real extreme-point symmetry;
if the component is a weighted-periodic function it degenerates to the odd-even type of extreme-point symmetry.
So ESMD\_III can give almost the same decomposition result as ESMD\_II for \emph{Example 3}.
By the way, for this case ESMD\_III only needs $5$ sifting times, which is much less than $29$ for ESMD\_II.

\vskip 1mm
It follows from Fig.18 that ESMD\_III also outputs good AGM curve for the temperature data. Its $11$ times sifting
yields a variance ratio $36.97\%$, which is slightly smaller than $37.11\%$ obtained by ESMD\_II with $29$ times
sifting. By making comparison with Fig.16, we see ESMD\_III yields less modes than that of ESMD\_II.
One reason is that, relative to the odd-even interpolation, the $3$-curve one gives much quicker
decreasing to the number of extreme points; Another reason is that, ESMD\_III adopts a lower type of symmetry.

\section{Direct Interpolating Approach for Instantaneous Frequency and Amplitude}

 Now that the period should be defined relative to a segment of time and
the frequency needs to be understood point by point, we conciliate this
 conflict by developing a ``direct interpolating (DI)" approach for it.
Just as analyzed in \emph{Section 1.4}, the instantaneous frequency should be capable of reflecting
the intermittent case in Fig.1, rather than excluding the adjacent-equal situation.
The detailed interpolation algorithm is as follows.

\subsection{Interpolation Algorithm}

\quad\enskip For the decomposed $n$ IMFs we calculate their instantaneous frequencies by implementing the following algorithm:

\vskip 1mm\noindent\textbf{Step 1}: For each IMF (denoted by $(t(k),y(k))$ with $1\leq k\leq N$) find all the interpolating points
which satisfy: $y(k)>y(k-1)\; \& \;y(k)\geq y(k+1)$, $y(k)\geq y(k-1) \;\& \;y(k)>y(k+1)$,
$y(k)<y(k-1) \;\& \;y(k)\leq y(k+1)$ or $y(k)\leq y(k-1)\; \& \;y(k)<y(k+1)$
and numerate them by $E_i(t_i, y_i)$ with $1\leq i\leq m$.

\vskip 1mm\noindent\textbf{Step 2}: If there are two adjacent $E_i$ such that $y_{i-1}=y_i$ or $y_i=y_{i+1}$, then define
the frequency interpolation coordinates by
$a_i=t(i), f_i=0$;
further if $E_i$ and $E_{i+1}$ are adjacent extreme points,
then define $a_{i-1}=(t_{i}+t_{i-2})/2, f_{i-1}=1/(t_{i}-t_{i-2})$
and $a_{i+2}=(t_{i+3}+t_{i+1})/2, f_{i+2}=1/(t_{i+3}-t_{i+1})$;
else if $E_i$ and $E_{i+1}$ (or $E_{i-1}$ and $E_i$) are not adjacent extreme points, then
define $a_{i-1}=t_{i-1}, f_{i-1}=1/[(t_{i+2}-t_{i-2})-(t_{i+1}-t_{i})]$
and $a_{i+2}=t_{i+2}, f_{i+2}=1/[(t_{i+3}-t_{i-1})-(t_{i+1}-t_{i})]$;
else, define $a_{i}=(t_{i+1}+t_{i-1})/2, f_{i}=1/(t_{i+1}-t_{i-1})$.

 \vskip 1mm\noindent\textbf{Step 3}: To add the boundary points with a linear interpolating method. For the left one,
  if it is an adjacent equal case, assign the value $f_1=0$ to $a_1=t(1)$; if not,
  assign the value $f_1=(f_3-f_2)(a_1-a_2)/(a_3-a_2)+f_2$ to $a_1=t(1)$; further if $f_1\leq 0$
  then assign $f_1=1/[2(t_2-t_1)]$ to $a_1=t(1)$.
   For the right one, if it is an adjacent equal case, assign the value $f_{m}=0$ to $a_{m}=t(N)$;
   if not, assign the value $f_m=(f_{m-1}-f_{m-2})(a_m-a_{m-1})/(a_{m-1}-a_{m-2})+f_{m-1}$ to $a_{m}=t(N)$;
   further if $f_m\leq 0$ then assign $f_m=1/[2(t_m-t_{m-1})]$ to $a_m=t(N)$.

 \vskip 1mm\noindent\textbf{Step 4}: To make the interpolation with all the discrete points
 $(a_i,f_i)$ and get a curve $f(t)$. To be meaningful,
we define the instantaneous frequency curve by $\max\{0,\;f(t)\}$.

\vskip 1mm\noindent\textbf{Step 5}: To output subplot frequency figures for all IMFs.

\vskip 1mm In addition, since the interpolating method for the instantaneous amplitude is much simpler,
we omit its algorithm.
For an IMF, its instantaneous amplitude curve
can be understood as the upper envelop interpolated by all the maxima points of this IMF
in the absolute-value form (rely on all the extreme points of the original one). In fact, for an IMF obtained under
the envelop-symmetry scheme or the odd-even extreme-point symmetry scheme,
its amplitude curve is almost equivalent to the upper envelop of the IMF itself
with the slow modulation.
But for the three-curve type the amplitude
may have too quick modulation which is not preferred.
By the way, to reflect the energy variation in a distinct way,
the instantaneous amplitude and frequency can be figured out together.

\subsection{Performance of DI Approach}

 In the following we test the
DI approach with the decomposition results given in Fig.12.
By executing the previous algorithm on the $4$ IMFs it yields
instantaneous frequencies and amplitudes in Fig.19.
It follows from F1 that there are several segments on which the
instantaneous frequency attains the Nyquist frequency $f_N=f/2$,
here $f=20$Hz stands for the sampling rate of the wind data.
For a given IMF, such as the third one, this kind of figure can
reflect clear variation of the frequency and amplitude. The comparison between F3 and A3 shows that, at $t=10$s,
the third IMF has a sharp oscillation with a very small amplitude.

Besides reflecting the time-variation of amplitude and frequency, the DI approach
can also provide an intuitional frequency distribution for the IMFs (see Fig.20).
It gives us a new understanding on the decomposition. \emph{Though, the neighboring two modes
may have the same frequency on the whole, this state rarely occurs at the same time. In another word,
the frequencies of these two neighboring modes do not overlap at the same time for almost all the case.
So the so-called ``frequency-overlapping puzzle" is not a real puzzle in the sense of time-variation.}
With this understanding, the decomposed modes can be seen as independent ones.
Each one represents an individual component which accords with a specific physical oscillation.
Certainly, it needs a mathematical theory to support the decomposition methodology.
\textsl{It is the second attractive topic of this paper left for mathematicians.}

\begin{figure}[!htbp]
\begin{minipage}[t]{0.45\linewidth}
\centerline{\hbox{\epsfig{figure=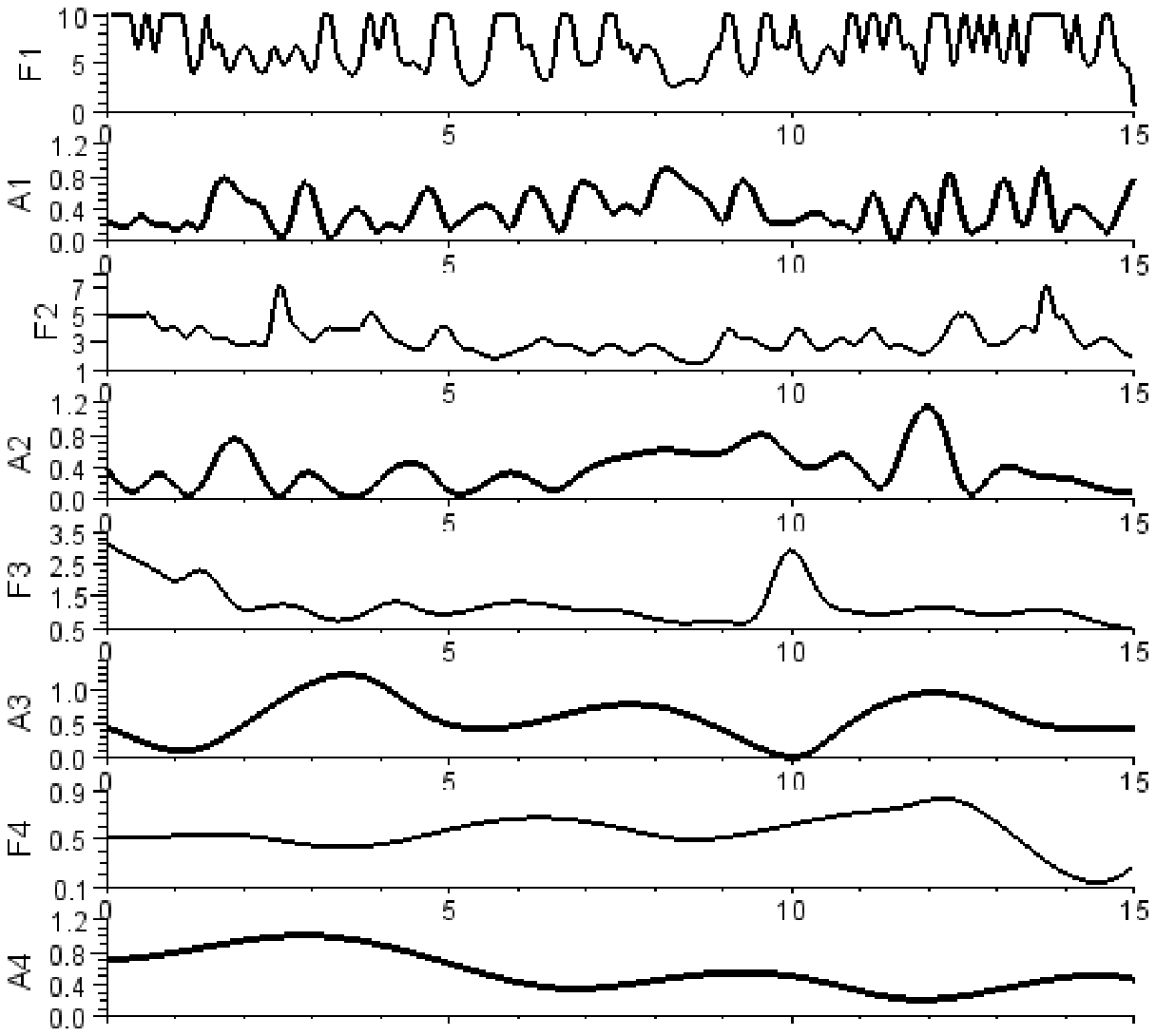,height=2.2in,width=2.5in,angle=0}}}
\caption{\footnotesize The instantaneous frequency (F) and amplitude (A) variations of the IMFs
for the wind data along the time. }
\end{minipage}
\hspace{4ex}
\begin{minipage}[t]{0.45\linewidth}
\centerline{\hbox{\epsfig{figure=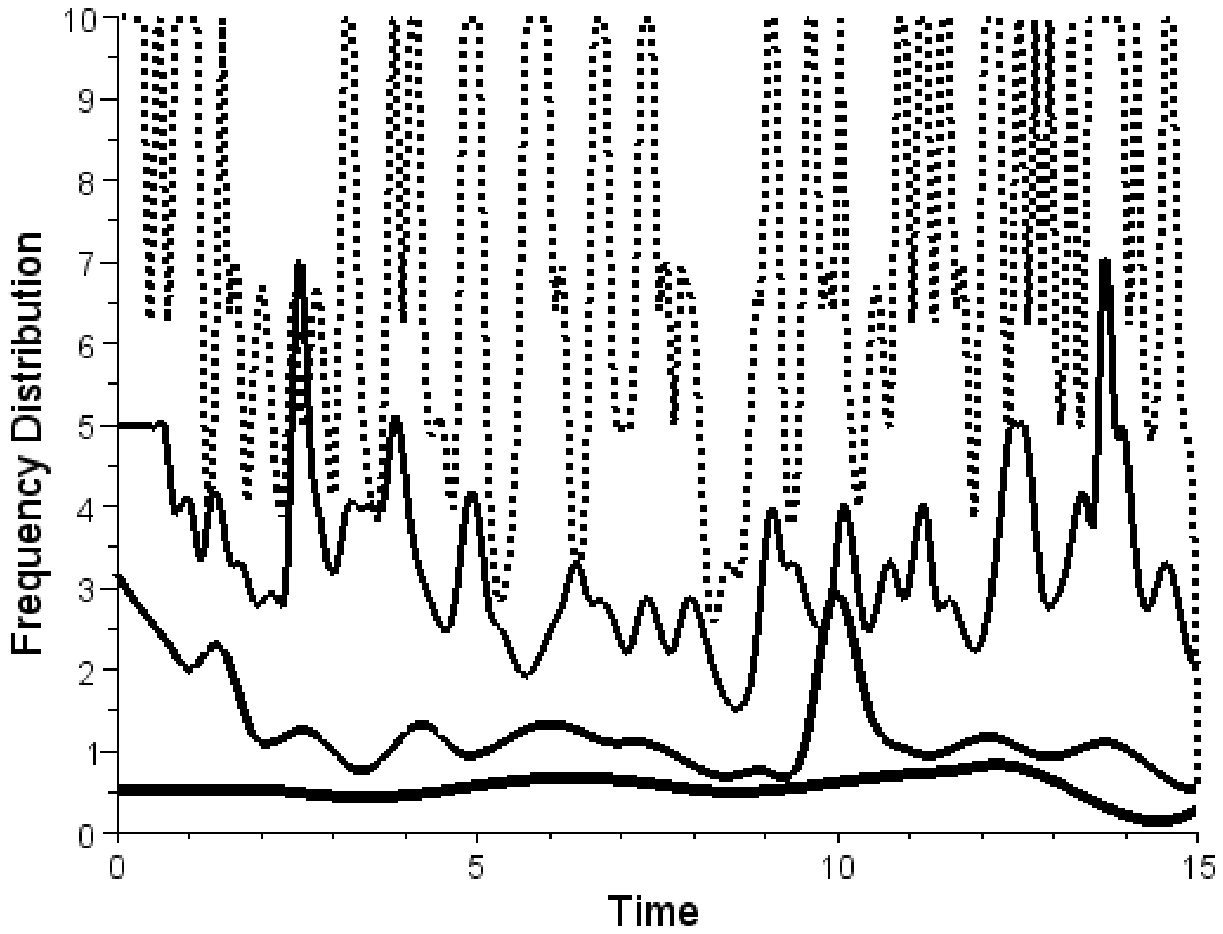,height=2.0in,width=2.5in,angle=0}}}
\caption{\footnotesize The frequency distribution of the IMFs for the wind data.}
\end{minipage}
\end{figure}

\vskip 1mm \emph{Since the frequency and amplitude (energy) of each IMF shift at any time, it is unreasonable
to project the total energy onto a series of frequencies as a Fourier
frequency-spectrum or a Hilbert time-frequency-spectrum,
 after all, the total energy itself changes along the time.}
 With this understanding we abandon the spectrum method and turn to
 discussing the time-variation of the total energy.
According to the state in \emph{Section 1.3}, the $j$-th IMF almost
accord with the mathematical expression $x_j(t)=A_j(t)\cos{\theta_j(t)}$ ($1\leq j\leq n$),
here $A_j(t)$ is actually the called amplitude curve.
Taking ESMD\_II as a default decomposition,
 the odd-even extreme-point symmetry scheme assigns
 $A_j(t)$ a slow modulation feather.
Based on this consideration, we define the so-called total energy
in the form of kinetic energy:
\begin{equation}
E(t)=\frac{1}{2}\sum\limits_{j=1}^n A^2_j(t).
\end{equation}
Certainly, here the word ``total energy" is in a general sense.
For the temperature data, it can be understood as
the whole oscillation intensity of the temperature.
By using this definition, we get the corresponding variation of the total energy for the wind data.
It follows from Fig.21 that the total energy of IMFs has three large peaks in the $15$s time segment.
By making comparison with Fig.22 we see
these peaks accord just right with the sunk parts of the
AGM. This is a very interesting phenomenon.
Perhaps it is caused by the energy transfer;
perhaps it is just a coincidence;
perhaps there are other deep causes.
\textsl{It is the third attractive topic of this paper left for further discussion.}

\begin{figure}[!htbp]
\begin{minipage}[t]{0.45\linewidth}
\centerline{\hbox{\epsfig{figure=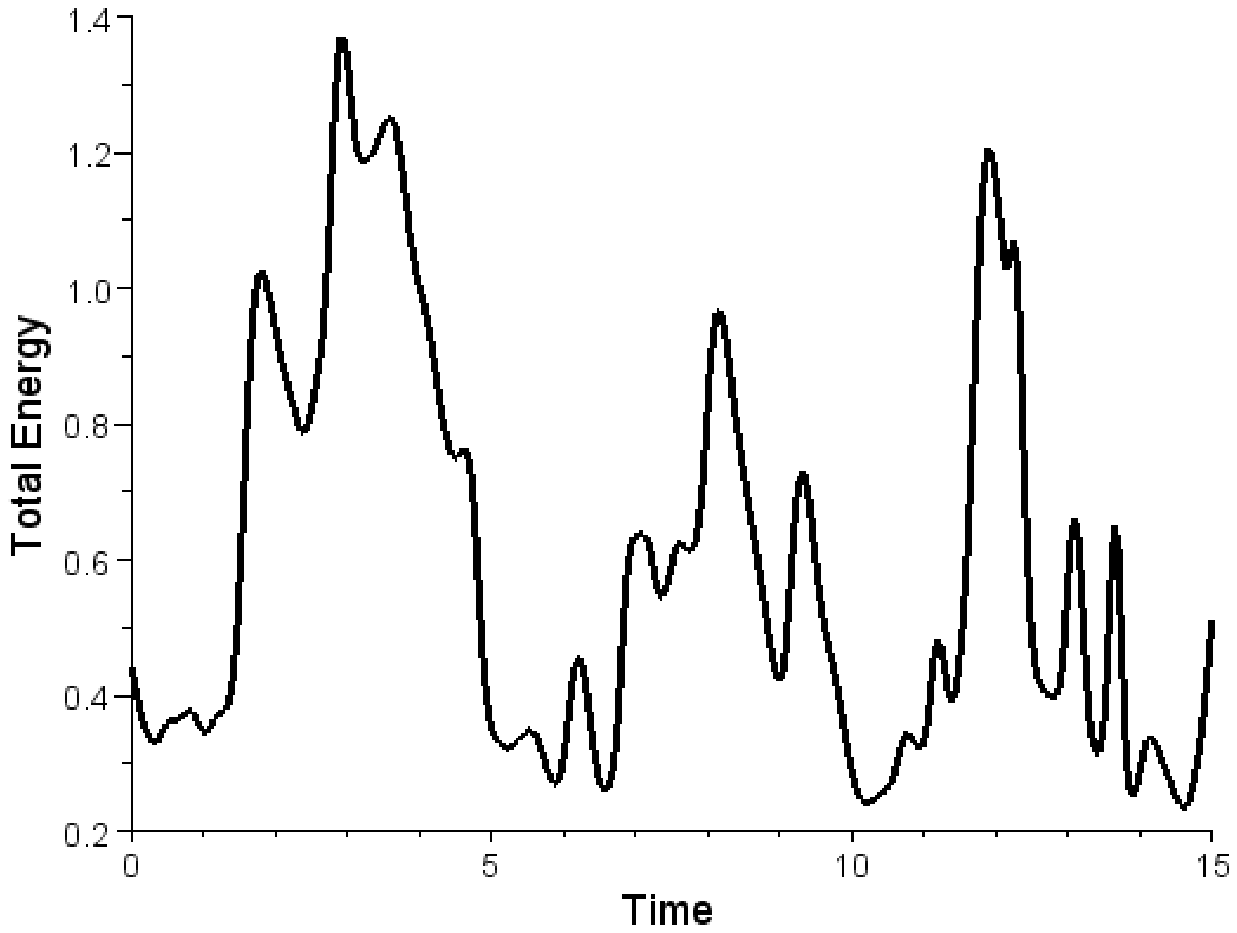,height=2.0in,width=2.3in,angle=0}}}
\caption{\footnotesize The time-variation of the total energy for all the IMFs of the wind data. }
\end{minipage}
\hspace{4ex}
\begin{minipage}[t]{0.45\linewidth}
\centerline{\hbox{\epsfig{figure=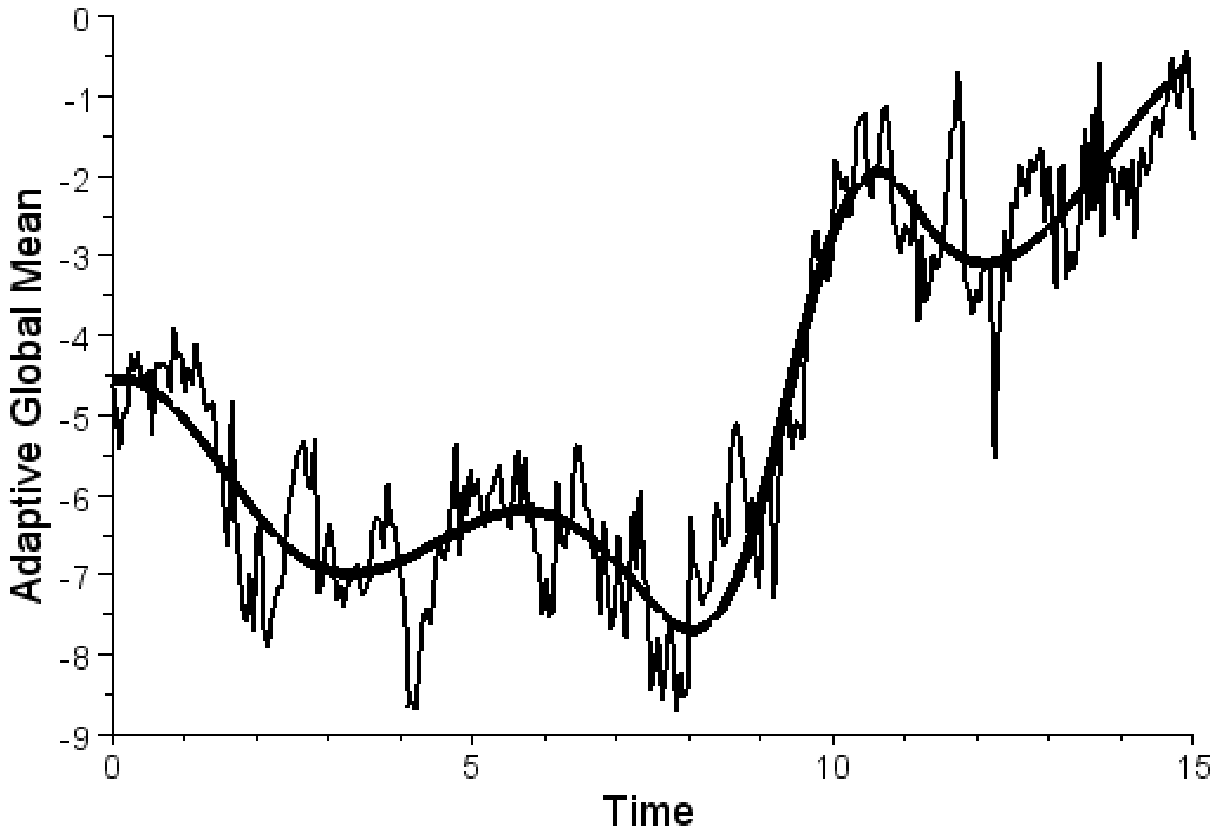,height=2.0in,width=2.3in,angle=0}}}
\caption{\footnotesize The variation of the optimal AGM curve for the wind data along the time.}
\end{minipage}
\end{figure}

\section{Summary}

 In order to analyze the non-stationary data, an ``extreme-point symmetric mode decomposition (ESMD)" method is proposed. There are two parts for it: the first part is the decomposition approach which yields a series of intrinsic mode functions (IMFs) together with an optimal ``adaptive global mean (AGM)" curve, the second part is the ``direct interpolating (DI)" approach which yields instantaneous amplitudes and frequencies for the IMFs together with a time-varying total energy. It can be seen as an alternate of the well-known ``Hilbert-Huang transform (HHT)" method with the following five characteristics:\\[1mm]
(1) Differing from constructing $2$ outer envelopes,
its sifting process is implemented by the aid of $1$, $2$, $3$ or more inner curves interpolated
by the midpoints of the line segments connecting the local maxima and minima points.
Accordingly, ESMD is classified into ESMD\_I, ESMD\_II, ESMD\_III, etc.\\[1mm]
(2) It does not decompose the data to the last trend curve with
at most one extreme point, it optimizes the residual component to be
an optimal AGM curve which possesses a certain number of extreme points.
By the optimizing process one can determine the sifting times and
output the optimal decomposition.\\[1mm]
(3) The concept of extreme-point symmetry is wider than the envelop symmetry.
 From the viewpoint of material movement, as a matter of fact, the oscillation
 occurs around the equilibrium which may also shift during this process.
 So the extreme-point symmetry actually reflects the local symmetry about itself.
ESMD\_I adopts a rigid extreme-point symmetry which is more rigorous than envelop symmetry;
 ESMD\_II adopts an odd-even type of extreme-point symmetry which
is almost equivalent to the envelop symmetry;
ESMD\_III adopts a three-curve type of extreme-point symmetry
which is more general than the envelop symmetry.\\[1mm]
(4) The definition of IMF is extended.
The new form not only includes the intermittent case
but also looses the requirement on symmetry.\\[1mm]
(5) The Hilbert-spectral-analysis approach for instantaneous frequency and amplitude
is substituted by the data-based one.
 This new approach easily conciliates the conflict:
 \emph{the period should be defined relative to a segment of time and
the frequency needs to be understood point by point.}
It is unreasonable to project the total energy onto a series of fixed frequencies as a Fourier
frequency-spectrum or a Hilbert time-frequency-spectrum because the total energy itself changes along the time.
The DI approach can not only yield clear distribution for instantaneous frequency and amplitude
but also reflect the time-variation of total energy in a distinct way.

\vskip 1mm Comparing ESMD\_I, ESMD\_II and ESMD\_III reveals that, as the number of interpolating curves increases,
1) the mode number decreases; 2) the symmetry degree decreases;
3) the amplitude modulation increases; 4) the decomposition efficiency increases
(needs less sifting times).
With these understandings, we prefer doing the decomposition with the eclectic one.
In fact, ESMD\_I can only yield acceptable imperfect decomposition
with low times sifting. The equal-amplitude phenomenon
will occur at high sifting times. In order to keep physical meaning this case should be avoided.
ESMD\_III has high decomposition efficiency,
but its low degree of symmetry and quick modulation of amplitude are disadvantageous for frequency and energy analysis.
The data processing tests also indicate that ESMD\_II is superior to
ESMD\_I and ESMD\_III and its decomposition result is more preferable.
In addition, from the first try on ESMD\_I to the successful exploitations of ESMD\_II and ESMD\_III it
costs us two years time.
Besides this paper, we have also developed a software which is now protected by the National Copyright Administration of China
[Wang and Li (2012), Wang and Li (2012b)].
ESMD\_II is anticipated an effective usage in the fields of atmospheric and oceanic sciences,
informatics, economics, ecology, medicine and seismology, etc.

\vskip 1mm Though these three types of interpolation have much difference,
they have a mutual advantage. All of them
can output good AGM curve. In fact, this advantage owes to the inner interpolation.
Besides the decomposition, the ESMD method also offers
a good adaptive approach for data fitting. It is superior to the common
least-square method and running mean approach.
In fact, the first one is awkward in application due to the request on a priori function form,
the second one lacks of theoretical basis and
different choices of time windows and weight coefficients may result in
different curves.

\vskip 1mm
We note that the type of interpolation is not essential,
 the approach developed here is also suitable for the envelop-symmetric scheme adopted by the EMD method
 [see \emph{Appendix B}].
As the stoppage criterion concerned, our strategy is an synthetic one.
A preset-error condition is adopted to ensure the symmetry and an ensemble optimal-sifting-times (OST) scheme
  is chosen to optimize the whole decomposition.
 In addition to this stoppage criterion there is also another applicable one.
  In fact, in view of the intermittent feature of IMF's symmetry [Wang and Li (2012a)], we can
 abandon the preset-error condition and execute the OST processing repeatedly and draw out a series of IMFs.
 If the mode symmetry is merely concerned this approach is a better choice,
 but it follows from the test in \emph{Appendix C} that the last residual may be an inferior one.
 In order to make up this defect, one can further implement the ensemble OST scheme on the whole IMF sets.
 Certainly, its time-consuming would be longer than the present one.
 The last appendix is about the asymptotic behavior of the modes along sifting times.

\vskip 4mm
\noindent\textbf{Acknowledgments.} Firstly, we  thank Professor Norden E. Huang for his
enthusiastic encouragement and many stimulating discussions on the topics related to the present research;
Secondly, we thank Professor Xian-Yao Chen for his help in the manuscript revision; Thirdly,
we thank the support from the Shandong Province Natural Science Fund, P.R. China
(No.ZR2012DM004).

\vskip 3mm
\noindent\textbf{Appendix A: Boundary Processing}
\vskip 2mm

In our program codes we have developed the linear interpolation method given by Wu and Huang (2009) and revised
the interpolation styles for the too steep boundary case.
This revision can make the boundary much more stable, even for the tests given by Wang and Li (2012a) with $100,000$
sifting times. Take the left boundary processing as an example.
Let $y(t)=k_1 t+b_1$ and $y(t)=k_2 t+b_2$ be the upper and lower lines interpolated by the first two
maxima and minima points, respectively. Also denote the first point of the data by $Y_1$.
According to Wu and Huang's classification,
(1) if $b_2\leq Y_1\leq b_1$, then define $b_1$ and $b_2$ as the boundary maximum and minimum points, respectively;
(2) if $Y_1>b_1$ (or $Y_1<b_2$), then define $Y_1$ and $b_2$ (or $b_1$ and $Y_1$) as the boundary maximum and minimum points.
 But if the boundary is too steep (relative to these two interpolating lines), this kind of processing may
 lead to instability to the decomposition. So we substitute the second term by:
 (2) if $b_1<Y_1\leq (b_1+b_2)/2+(b_1-b_2)=(3b_1-b_2)/2$ (or
 $(3b_2-b_1)/2=(b_1+b_2)/2-(b_1-b_2)\leq Y_1<b_2$), then define $Y_1$ and $b_2$ (or $b_1$ and $Y_1$) as the boundary maximum and minimum points;
 (3) if $Y_1>(3b_1-b_2)/2$ (or $Y_1<(3b_2-b_1)/2$), then define $Y_1$ as the boundary maximum (minimum) point
 and take the boundary minimum (maximum) point by using new interpolating line from the
 first minimum (maximum) point: $y(t)=k^* t+b^*$, here $k^*$ relies on the point $(0, Y_1)$ and the
 first maximum (minimum) point. The detailed processing is depicted in Fig.23.
\begin{figure}[!htbp]
\centerline{\hbox{\epsfig{figure=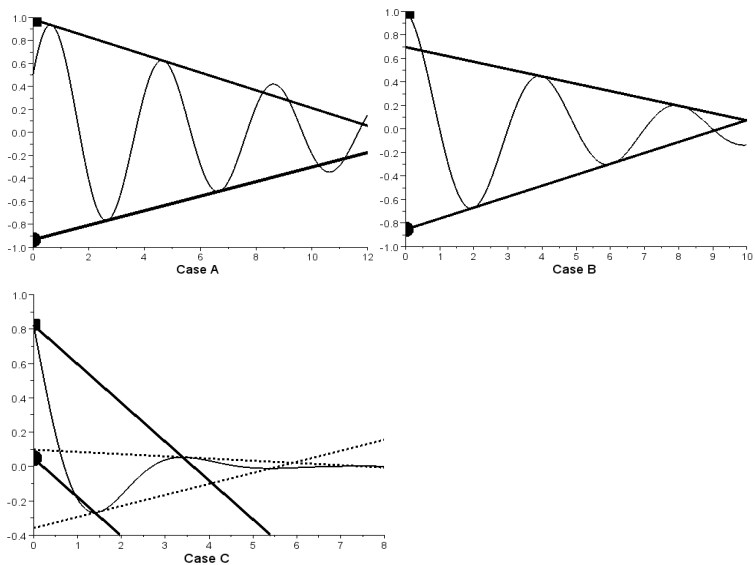,height=3in,width=4in,angle=0}}}
\caption{\footnotesize The developed linear interpolation method for the left boundary processing.
Case A: $b_2\leq Y_1\leq b_1$; Case B: $b_1<Y_1\leq (3b_1-b_2)/2$;
Case C: $Y_1>(3b_1-b_2)/2$. }
\end{figure}

\noindent\textbf{Appendix B: Decomposition with Envelop-Symmetric Scheme}
\vskip 2mm

The approach developed in this paper is also suitable for the envelop-symmetric scheme adopted by the EMD method.
At this time, the decomposition result is similar to that of odd-even extreme-point symmetric one.
To make comparison with ESMD\_II at $30$ sifting times, we choose the second optimal one $39$ in the interval $[1,100]$
(the first optimal one is $62$).
For this case the AGM curve is as good as ESMD\_II with variance ratio $\nu=33.8\%$
and Mode1 and Mode2 are very similar to that of Fig.12.
The difference lies in the low-frequency modes (Mode3 and Mode4).
This indicates the whole outer symmetry is similar to the local inner symmetry
for the case with dense extreme points.
But for the sparse case, the outer and inner interpolations may result in
different results.
Notice that the magnitudes of the upper and lower outer envelopes
are bigger than that of odd and even inner curves,
their interpolating uncertainty should be bigger than the inner ones for the sparse case.
Hence, the result given by ESMD\_II should be more credible.
\begin{figure}[!htbp]
\begin{minipage}[t]{0.45\linewidth}
\centerline{\hbox{\epsfig{figure=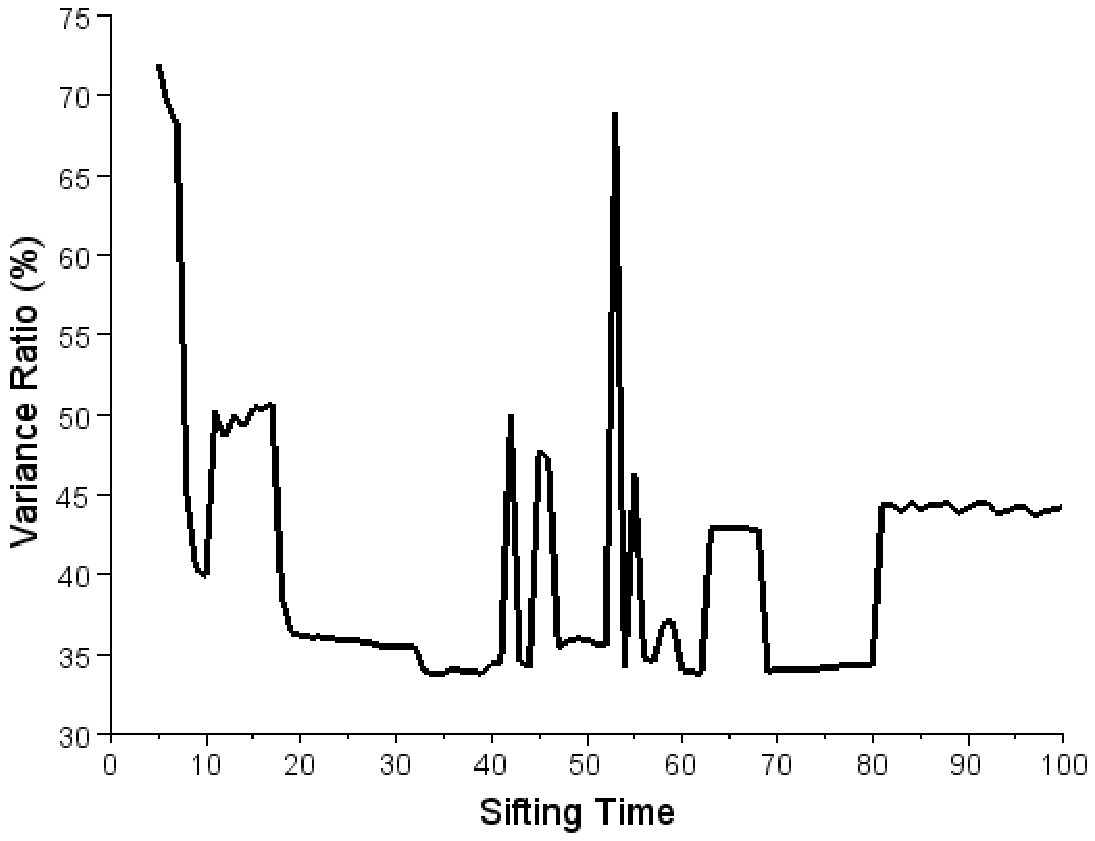,height=1.8in,width=2.3in,angle=0}}}
\caption{\footnotesize Under the envelop-symmetric scheme, the distribution of the variance ratio
$\nu=\sigma/\sigma_0$ along the sifting times for the wind data. }
\end{minipage}
\hspace{3ex}
\begin{minipage}[t]{0.45\linewidth}
\centerline{\hbox{\epsfig{figure=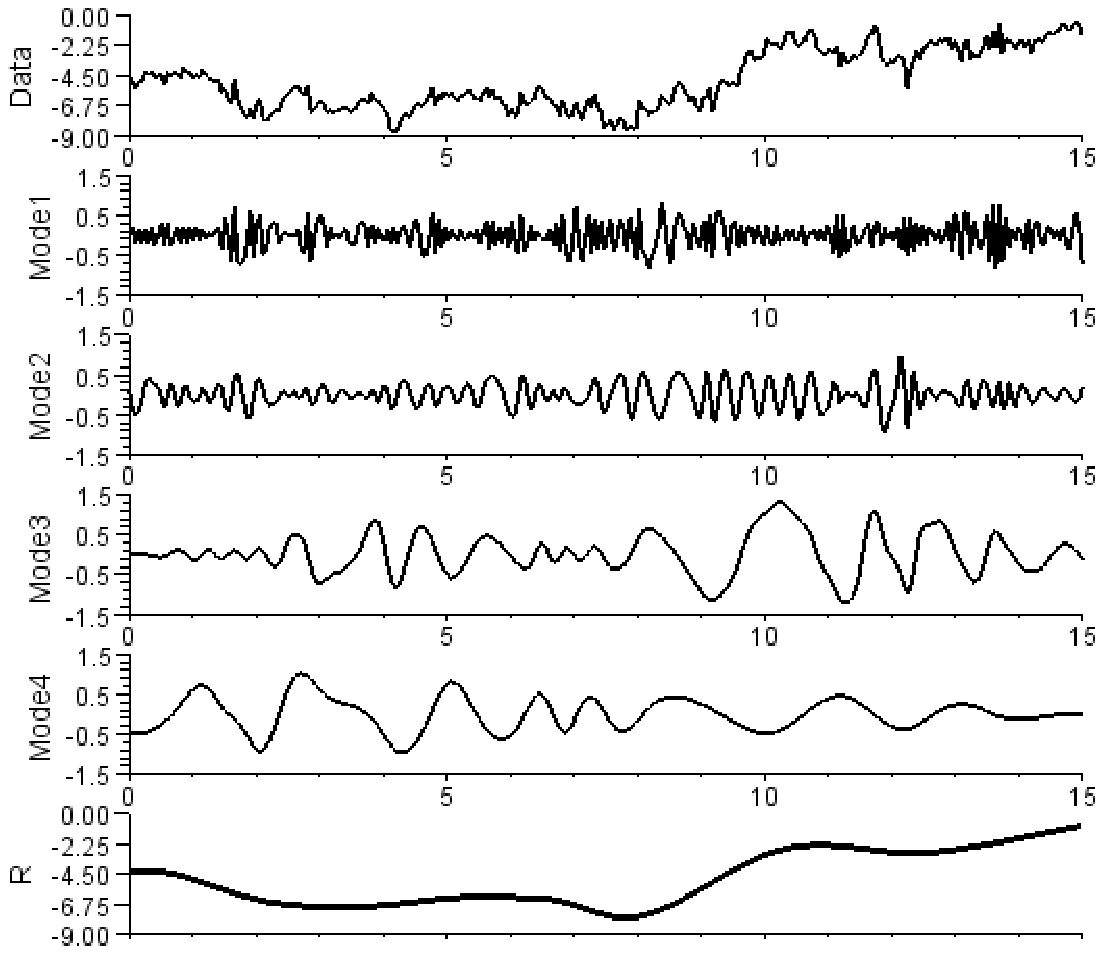,height=2.1in,width=2.3in,angle=0}}}
\caption{\footnotesize Under the envelop-symmetric scheme, the decomposition result
for the wind data with $39$ sifting times, here
the horizontal axis stands for the time (second). }
\end{minipage}
\end{figure}

\noindent\textbf{Appendix C: Decomposition with Optimal-Sifting-Times Approach}
\vskip 2mm

 In the following we try another stoppage criterion which is associated with the single usage of
optimal-sifting-times (OST) approach.
For a preset maximum sifting times $K_{max}$,
we can select an optimal one in the integer interval $[1, K_{max}]$
which accords with the minimum value of $A_{max}$,
where $A_{max}$ stands for the maximum amplitude of an quasi-mode's mean cure $L^*$.
By implementing this OST approach repeatedly we can draw out a series of IMFs.
For convenience of processing we take the envelop-symmetric scheme as an example.
At this time, $L^*$ is simply the mean of the upper and lower envelops and the corresponding
decomposition results are given in Fig.26 and 27 with $K_{max}=100$ and $200$, respectively.
Though in the interval $[1, 200]$ we have more choices and the symmetry of IMFs should be better,
its last residual R is worse than that in $[1, 100]$. That is to say,
the optimal processing on IMFs can not ensure an optimal global mean cure.
This defect can be made up by further optimizing $K_{max}$
with respect to the variance ratio $\nu=\sigma/\sigma_0$. Certainly, it would cost a longer
time-consuming than the one adopted by this paper due to more times of calculation.

\begin{figure}[!htbp]
\begin{minipage}[t]{0.45\linewidth}
\centerline{\hbox{\epsfig{figure=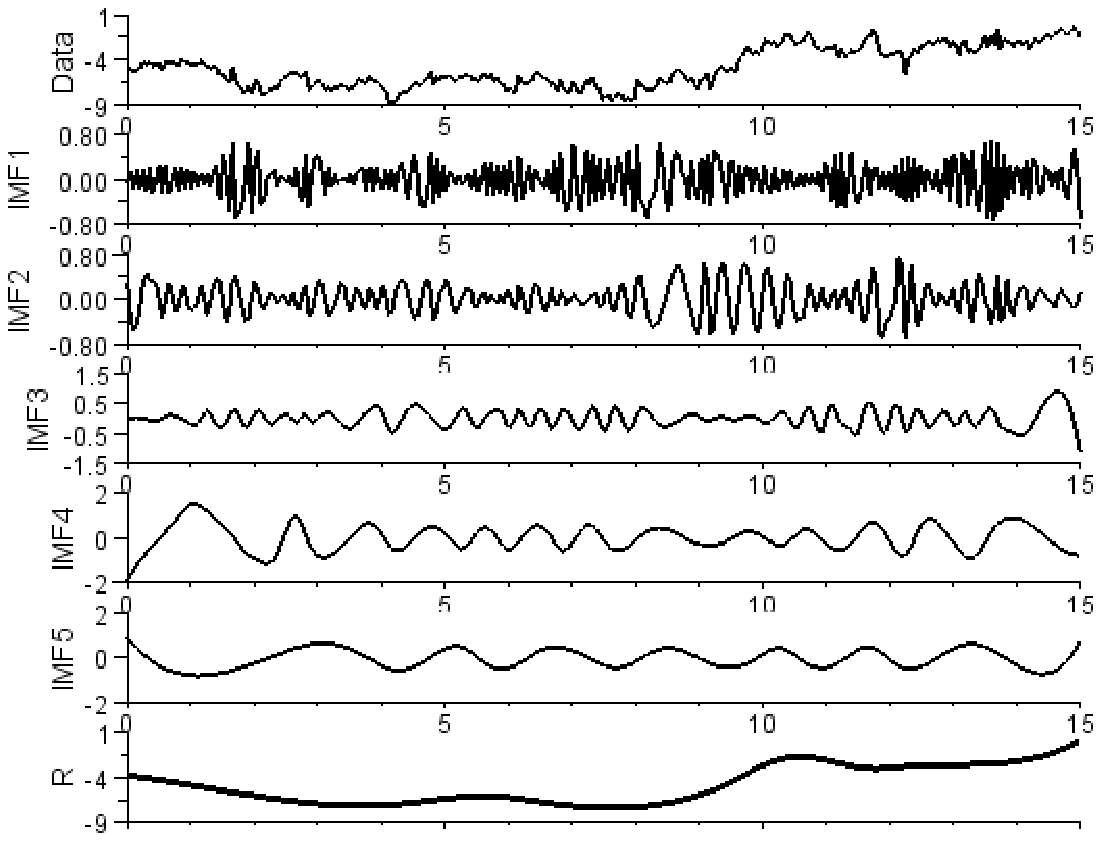,height=2.1in,width=2.3in,angle=0}}}
\caption{\footnotesize Under the envelop-symmetric scheme, the decomposition result
for the wind data given by optimal-sifting-times (OST) approach with $K_{{max}}=100$, here
the horizontal axis stands for the time (second). }
\end{minipage}
\hspace{3ex}
\begin{minipage}[t]{0.45\linewidth}
\centerline{\hbox{\epsfig{figure=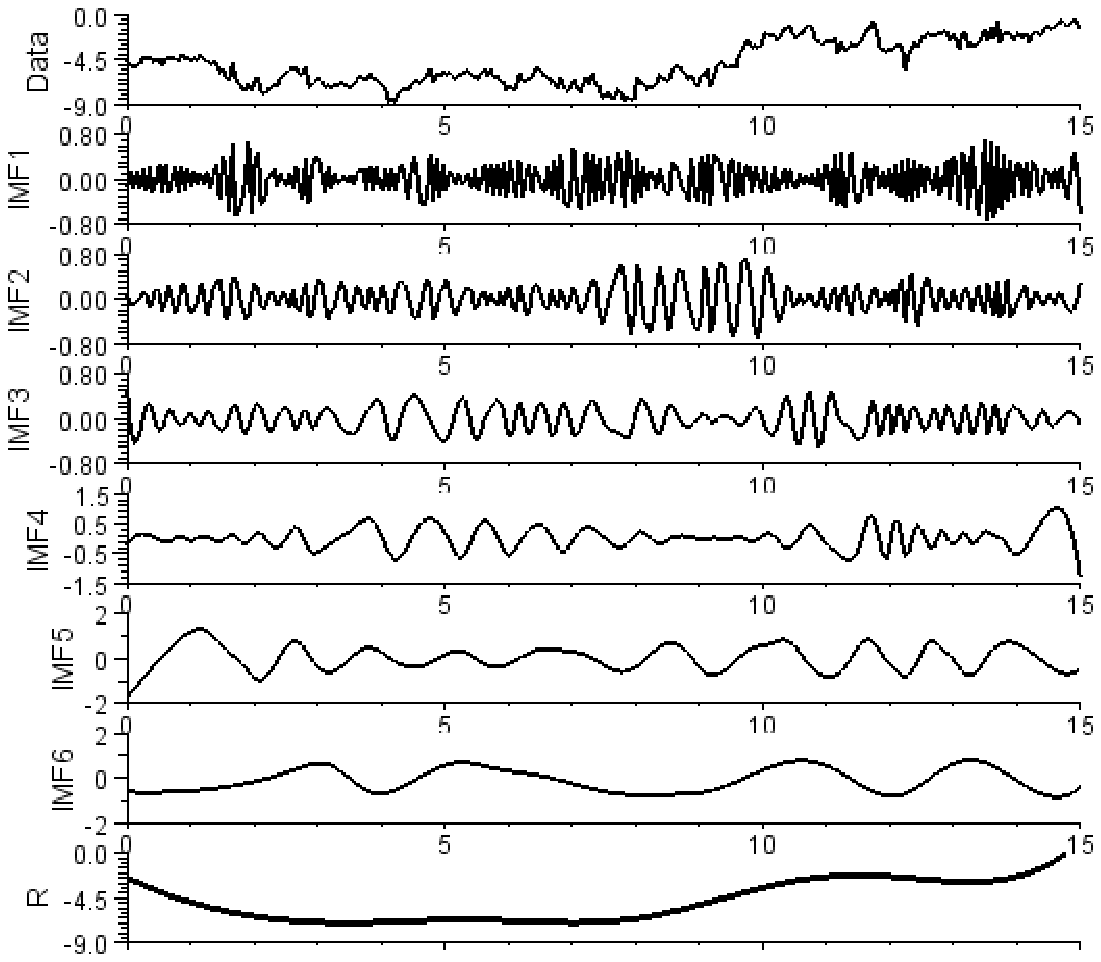,height=2.1in,width=2.3in,angle=0}}}
\caption{\footnotesize Under the envelop-symmetric scheme, the decomposition result
for the wind data given by optimal-sifting-times (OST) approach with $K_{{max}}=200$, here
the horizontal axis stands for the time (second). }
\end{minipage}
\end{figure}

\noindent\textbf{Appendix D: Asymptotic Behavior of the Modes}
\vskip 2mm

 We note that the degree of symmetry does not increase uniformly along the sifting times.
Just as indicated by Wang and Li (2012a), its variation behaves in an intermittent manner [see Fig.28].
In addition to it, there is another phenomenon that, by and large,
the number of extreme-points for the first mode increases with the sifting times.
The ultimate case for it is that all the extreme-points are adjacent with no middle points.
At this time, the instantaneous frequency attains its maximum value (equals to the Nyquist frequency).
Hence, in order to revealing the intrinsic oscillation of a process through mode decomposition,
too high times sifting is not recommended. From our experiential understanding, dozens of times
sifting is enough for the ESMD decomposition.
\begin{figure}[!htbp]
\centerline{\hbox{\epsfig{figure=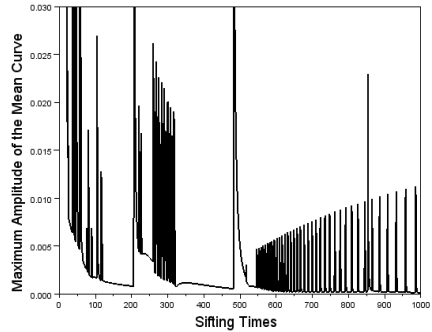,height=1.8in,width=2.3in,angle=0}}}
\caption{\footnotesize The variation of maximum amplitude for the mean curve of IMF1
along the sifting times, here the wind data is used.}
\end{figure}



\begin{thebibliography}{00}
\bibitem{s1} Bao, S. et al. (2011) An empirical study of tropical cyclone activity in the Atlantic and Pacific oceans:1851-2005.
\textit{Adv. in Adap. Data Analy.}, \textbf{3}(3): 291每307.

\bibitem{s2} Flandrin, P., Rilling, G. and Goncalves, P. (2004). Empirical mode decomposition as a
filterbank.\textit{ IEEE Data Proc. Lett.}, \textbf{11}: 112-114.

\bibitem{s3} Flandrin, P. and Goncalves, P. (2004). Empirical mode decompositions as datadriven
wavelet-like expansions. \textit{Int. J. Wavelets, Multiresolut. Inf. Process.}, \textbf{2}(4):
477-496.

\bibitem{s4} Flandrin, P., Goncalves, P. and Rilling, G. (2005). EMD equivalent filter bank, from
interpretation to applications. in \textit{Hilbert-Huang Transform and Its Applications}, Eds.
N. E. Huang and S. Shen, World Scientific, Singapore, pp. 57-74.

\bibitem{s5} Hou, T. Y., Yan, M. P. and Wu, Z. (2009). A variant of the EMD method for multiscale
data. \textit{Adv. in Adap. Data Analy.},  \textbf{1}(4): 483每516.

\bibitem{s6} Hou T. Y. and Shi Z. Q. (2011). Adaptive data analysis via sparse time-frequency represen
tation. \textit{Adv. in Adap. Data Analy.}, \textbf{3}(1,2): 1-28.

\bibitem{s7} Hou T. Y. and Shi Z. Q. (2012). Data-driven time-frequency analysis.
arXiv:1202.5621v1 [math.NA].


\bibitem{s8} Huang, N. E. et al. (1998). The empirical mode decomposition and the Hilbert spectrum
for nonlinear and non-stationary time series analysis. \textit{Proc. R. Soc. Lond. A}, \textbf{454}:
903每995.

\bibitem{s9}  Huang, N. E. et al. (2003). A confidence limit for the empirical mode decomposition and
Hilbert spectral analysis. \textit{Proc. R. Soc. Lond. A}, \textbf{459}: 3217每2345.

\bibitem{s10}  Huang, N. E. and Shen, S. S. P. (Ed.)(2005). \textit{Hilbert-Huang Transform: Introduction and Applications},
World Scientific, Singapore, 311pp.

\bibitem{s11} Huang, N. E. and Wu, Z. (2008). A review on Hilbert-Huang transform: method and its
applications to geophysical studies. \textit{Rev. Geophys.}, \textbf{46}(2): RG2006.

\bibitem{s12} Huang, N. E. et al. (2009a). On instantaneous frequency. \textit{Adv. in Adap. Data Analy.},
\textbf{1}(2): 177-229.

\bibitem{s13} Huang, N. E. et al. (2009b). Reductions of noise and uncertainty in annual global surface
temperature anomaly data. \textit{Adv. in Adap. Data Analy.}, \textbf{1}(3): 447每460.

\bibitem{s14} Moghtaderi, A., Borgnat, P. and Flandrin P. (2011).
 Trend filtering: empirical mode decompositions versus L1 and Hodrick-Prescott.
 \textit{Adv. in Adap. Data Analy.}, \textbf{3}(1,2): 41每61.

\bibitem{s15} Moghtaderi, A., Flandrin P. and Borgnat, P. (2011). Trend filtering via empirical mode decompositions.
\textit{Computational Statistics and Data Analysis},
doi:10.1016/ j.csda.2011.05.015.

\bibitem{s16} Rilling, G., Flandrin, P. and Goncalves, P. (2003). On empirical mode decomposition and
its algorithms. \textit{IEEE-EURASIP Workshop on Nonlinear Data and Image Processing
NSIP-03}, Grado (I).

\bibitem{s17} Rilling, G. and Flandrin, P. (2008). One or two frequencies? The empirical mode decomposition answers.
\textit{IEEE Trans. on Data Processing}, \textbf{56} (1): 85-95.

\bibitem{s18} Wang, G. et al. (2010). On intrinsic mode function. \textit{Adv. in Adap. Data Analy.},
\textbf{2}(3): 277每293.

\bibitem{s19} Wang, J.L. and Li, H. F. (2006). The weighted periodic function and its properties.
\textit{Dynamics of Continuous Discrete and Impulsive Systems}, \textbf{13}(S3): 1179-1183.

\bibitem{s20} Wang, J.L. and Zhang, G. (2006). Asymptotic weighted periodicity for delay differential equations.
\textit{Dynamic Systems and Applications}, \textbf{15}: 479-500.

\bibitem{s21} Wang, J.L. and Li, H. F. (2007). Asymptotic weighted-periodicity of the impulsive parabolic equation with time delay. \textit{Acta Mathematicae Applicatae Sinica}, \textbf{23}(1): 1-8.

\bibitem{s22} Wang, J.L. and Li, H. F. (2011). Surpassing the fractional derivative: Concept of the memory-dependent derivative. \textit{Computers and Mathematics with Applications}, \textbf{62}: 1562-1567.

\bibitem{s23} Wang, J.L. and Li, H. F. (2012). Software for the direct interpolating method to the
frequency under the frame of extreme-point symmetric mode decomposition method.
 Computer Software Copyright Registration, No.2012SR102181, from the National Copyright Administration of China.

\bibitem{s24} Wang, J.L. and Li, Z. J. (2012a). What about the asymptotic behavior of the intrinsic
mode functions as the sifting times tend to infinity? \textit{Adv. in Adap. Data Analy.},
\textbf{4}(1,2): 1250008 (1-17).

\bibitem{s25} Wang, J.L. and Li, Z. J. (2012b). Software for the extreme-point symmetric mode decomposition method to
nonlinear and non-stationary signal processing. Computer Software Copyright Registration, No.2012SR052512, from
the National Copyright Administration of China.

\bibitem{s26} Wu, Z. and Huang, N. E. (2005). Statistical significant test of intrinsic mode functions.
In \textit{Hilbert-Huang Transform: Introduction and Applications}, pp. 125每148, Eds. N. E.
Huang and S. S. P. Shen, World Scientific, Singapore, 311pp.

\bibitem{s27} Wu, Z. and Huang, N. E. (2009). Ensemble empirical mode decomposition: a noise
assisted data analysis method. \textit{Adv. in Adap. Data Analy.}, \textbf{1}(1): 1每41.

\bibitem{s28} Wu, Z. and Huang, N. E. (2010). On the filtering properties of the empirical mode decomposition.
\textit{Adv. in Adap. Data Analy.}, \textbf{2}(4): 397-414.

\bibitem{s29} Wu, H. T., Flandrin P. and Daubechies I. (2011). One or two frequencies? the synchrosqueezing answers.
\textit{Adv. in Adap. Data Analy.}, \textbf{3}(1,2): 29-39.

\end{thebibliography}
\end{document}